\documentstyle[amsmath,amssymb,graphicx]{article}

\def\be{\begin{eqnarray}}
\def\ee{\end{eqnarray}}
\def\nn{\nonumber}

\def\p{\partial}

\hoffset= - 1.0in         

\title{{\bf On AGT relation in the case of $U(3)$
} \vspace{.2cm}}
\author{{\bf A.Mironov}\footnote{ {\small {\it
Lebedev Physics Institute} and {\it ITEP, Moscow, Russia}};
mironov@itep.ru; mironov@lpi.ru} \ and {\bf
A.Morozov}\thanks{{\small {\it ITEP, Moscow, Russia}};
morozov@itep.ru} \date{ }}

\begin{document}

\maketitle

\vspace{-5.0cm}

\begin{center}
\hfill FIAN/TD-16/09\\
\hfill ITEP/TH-32/09
\end{center}

\vspace{3.cm}

\begin{abstract}
\noindent We consider the AGT relation \cite{AGT}, expressing
conformal blocks for the Virasoro and W-algebras in terms of
Nekrasov's special functions, in the simplest case of the 4-point
functions for the first non-trivial $W_3$ algebra. The standard set
of Nekrasov functions is sufficient only if additional null-vector
restriction is imposed on a half of the external $W$-primaries and
this is just the case when the conformal blocks are fully dictated
by $W$-symmetry and do not depend on a particular model. Explicit
checks confirm that the AGT relation survives in this restricted
case, as expected.
\end{abstract}

\bigskip

\bigskip

\tableofcontents

\section{Introduction}

A recent paper \cite{AGT}, see also \cite{Wyl}-\cite{mmNF},
provided a
long-anticipated relation between $2d$ conformal theories \cite{BPZ}
and Nekrasov's functions
\cite{Nek}, decomposing the Seiberg-Witten prepotential into sums
over $4d$ instantons. An exact formulation of the AGT relation \cite{AGT}
is that the {\bf universal conformal blocks \cite{BPZ,Za,ZZ} for
generic
Verma modules of the Virasoro algebra coincide with
the Nekrasov partition functions, provided
the $\alpha$-parameters of the Virasoro primary fields are
linear combinations of the
$a$ and $\mu$-parameters of the Nekrasov functions.}
The central charge is related to Nekrasov's
peculiar additional parameter $\epsilon=\epsilon_1+\epsilon_2$,
$c = 1 + \frac{6\epsilon^2}{\epsilon_1\epsilon_2}$ and
dimensions are given by
$\Delta_\alpha = \frac{\alpha(\epsilon-\alpha)}{\epsilon_1\epsilon_2}$.
For details of explicit check of this relation see
\cite{AGT,Wyl,MMMagt}.
An additional observation of \cite{AGT} is that with the same
$\{\alpha\} - \{a,\mu\}$ identification of the perturbative part of
the Nekrasov partition function (exponentiated
perturbative part of the Seiberg-Witten prepotential)
coincides with the DOZZ functions \cite{DOZZ}: the structure
constants of concrete, Liouville conformal model, that is,
a certain reduction of the $SU(2)$ WZNW model,
which has a natural description in terms of a single $2d$
free field \cite{fref}. With no surprise Nekrasov's functions,
appearing in this original AGT relation are associated with
the gauge group $SU(2)$.

An obvious generalization \cite{Wyl} of the AGT relation
is to take a $2d$ conformal model made from the $r=N-1$ free fields
with the extended $W_N$-symmetry \cite{ZW}-\cite{FLit} and look if
its conformal blocks are expanded into the $SU(N)$ Nekrasov functions.

The first problem here is that the theory of W-symmetries is
under-developed. In the Virasoro sector, the crucial idea is to
identify Virasoro operators with the $2d$ stress-tensor, which
generates shifts in the holomorphic $z$ coordinate. This
identification results in the two well-known postulates: \be
L_{-1}V(z) = \p_zV(z) \label{L-1} \ee for arbitrary vertex operator
$V(z)$ and \be <V_{\Delta_1}(z_1) V_{\Delta_2}(z_2)
V_{\Delta_3}(z_3)> \ \sim\ z_{12}^{\Delta_3-\Delta_1-\Delta_2}
z_{23}^{\Delta_1-\Delta_2-\Delta_3}
z_{31}^{\Delta_2-\Delta_1-\Delta_3} \label{3pfV} \ee for an
arbitrary spherical 3-point function of the Virasoro {\it
quasi-primaries} \cite{BPZ}. The 3-point correlator of secondary
fields can be then immediately constructed from this one by
exploiting the Virasoro symmetry. After that, arbitrary conformal
blocks and correlators can be built from these three point functions
by a "gluing-of-pants" procedure, at least in principle
\cite{BPZ,Sonoda,MS}. Though it is a difficult procedure, involving
infinite summations, it is unambiguously defined. In the case of
$W$-symmetries, there is still no clear choice for the
$W$-generalization of $z$ (a modulus of the $W$-geometry, see, e.g.,
\cite{Wgeom}) and, hence, no formulas like (\ref{L-1}) and
(\ref{3pfV}). In result, an arbitrary theory with $W$-symmetry has
many additional free parameters: the entire set of the $3$-point
functions in the $W_3$ case is \be \ <V_{\vec\alpha_1}(1)\ \
V_{\vec\alpha_2}(\infty)\ \ (W_{-1})^kV_{\vec\alpha}(0)>
\label{omegadef} \ee involving a single insertion of arbitrary power
of $W_{-1}$. In a theory with underlying $W$-geometry, like the conformal Toda
proposed as an obvious $SU(N)$-counterpart of the Liouville theory
in \cite{Wyl}, these infinitely many parameters are of course fixed,
but the relevant theory is still unavailable.

Another obvious problem concerns the AGT relations themselves.
For $N>2$ there is a clear mismatch in the number of parameters.
Each $W_N$ primary, external or internal, is labeled by
the highest-weight, which is arbitrary $r=N-1$ dimensional vector
$\vec\alpha$, thus the 4-point conformal block depends on
$(4+1)\cdot(N-1) + 1 = 5N-4$ free parameters ($4+1$ is the number
of external + internal states, an extra $1$ is the central charge).
At the same time, Nekrasov function depends on $N-1$
"expectation values" $\vec a$, $2N$ "masses" $\mu_f$ and
two $\epsilon$-parameters, what gives $(N-1)+2N+2-1 = 3N$
($1$ is subtracted because conformal invariance forces
Nekrasov functions to be dimensionless:
to depend only on ratios of parameters).
For $N=2$ there was a nice matching \cite{MMMagt}: $5N-4$ = $3N$,
but it breaks down for $N>2$: some $2(N-2)$ parameters of
the conformal block do not have immediate counterparts in the
Nekrasov functions.

These two difficulties could be considered as a strong appeal for
developing the theory of $W$-symmetries and looking for
a further generalization of Nekrasov functions.
However, Niclas Wyllard suggested in \cite{Wyl} an elegant
way out which overcomes both problems at once.
One can restrict consideration
to the case when the appropriate number
of "external" states are the {\it special} ones \cite{FLit},
satisfying additional $W$-null-vector constraints at level one\footnote{
In fact, restriction to the special states is
rather natural within Davide Gaiotto's construction \cite{Gpr}
behind the AGT relation, where $4d$ SYM theory is described
as a result of 5-brane compactification on a Riemann
surface (spectral curve) {\it a la} \cite{Wbr,C}.
For $N\geq 3$ the punctures on the Riemann surface are not
of general type, and it was suggested in \cite{Wyl}
to identify them exactly with the {\it special} states
on the CFT side of the AGT relation.
We are grateful to the referee of the paper for his comments on this point.
In the present text, we discuss a possibility to express
conformal blocks through Nekrasov functions {\it without}
explicit reference to branes or to instanton expansions in
Seiberg-Witten theory, i.e. we are using AGT
relations as a tool to find explicit expressions for
conformal blocks, and then restriction to the {\it special}
states should be eliminated one day, perhaps, by further
extending the set of Nekrasov functions.
}.
In the case of $W_3$ just one is enough,
\be
(3w L_{-1} - 2 \Delta W_{-1})V_{\Delta,w} = 0,
\label{Wzc}
\ee
for higher $N$ the number of constraints
grows up to $N-2$.\footnote{
To avoid possible confusion, note that in \cite[(2.12)]{FLit}
some three constraints were imposed on one of the special states.
This was, however, done to make {\it a particular
method} working, as to the conformal block, it is well defined
(but very hard to evaluate) already with a single constraint
for the special state.}
Such "special" primary states form a one-parametric family
instead of an $r$-parametric one.
"The appropriate number" to solve the matching problem is
two out of four external states for the
$4$-point conformal block.
At the same time, if one of the three primaries
in (\ref{omegadef}) is special, these $3$-point functions
get unambiguously defined without any reference to
$W$-geometry. Thus, taking two of external states
to be {\it special}, one also eliminates the first problem.
Moreover, just under this restriction
a counterpart of the DOZZ functions for the $SU(3)$ conformal
Toda has been evaluated in \cite{FLit}, and \cite{Wyl}
checks that they indeed coincide
with the exponential of perturbative $SU(3)$ prepotential
in the conformal invariant case $N_f=2N=6$.
In particular, the linear relation between
$\alpha$ and $a,\mu$ parameters was established in this case.

Unfortunately no detailed
check of the main part of the AGT relation,
between conformal blocks and "instanton sums",
was explicitly made in \cite{Wyl}, despite relevant
formulas were extracted from \cite{ZW}-\cite{FLit}.
It is our task in this paper to partly fill in this gap.
We explicitly present this check in the  simplest case.
The "simplest" involves the following restrictions:
only $N=3$ and only
contributions at levels one and two are considered.
It is not a big problem to lift any of these two restrictions,
still calculations are tedious, and it is unclear if
they are that necessary to provide even
more evidence in support of the AGT relation.
In any case, they remain for the future work, maybe they can help to
better understand the notion of $W$-symmetry beyond
the seemingly-artificial (\ref{Wzc}) constraint and
to reveal the true meaning of the AGT relations,
and the way to prove it in the full generality.
This would further promote Nekrasov functions in the role
of the crucially important new special functions and stimulate
the further study of
matrix-model tau-functions (i.e. generalized $\tau$-functions
\cite{gentau}, subjected to additional Virasoro-like constraints),
already proposed for this role in \cite{amm}.

\bigskip

Presentation in this paper does {\it not} follow literally the
by-now-standard one, accepted in \cite{AGT,Wyl,MMMagt}.
Instead of just writing down the conformal blocks and Nekrasov
functions and comparing them (actually reporting the result of
computer calculations), we {\it also} try to demonstrate that
the AGT relation itself can turn into a powerful alternative
method of iterative, level-by-level, evaluation of conformal blocks.
Therefore, we begin with making anzatze for the
elementary constituents of conformal blocks
(the $2-$ and $3-$point functions)
and then define the remaining coefficients by the requirement
that the result matches Nekrasov functions.
A part of this comparison is to check that the Kac determinant
(that of the Shapovalov matrix made from the scalar products
of states, essentially a $2$-point functions at special points
$z_1=0$, $z_2=\infty$)) factorizes in an appropriate way:
in $\alpha$-parametrization of primary fields
its roots are given by linear function
$\vec\alpha\vec e_i = m_i \epsilon_1 +n_i \epsilon_2$.
The AGT relation strengthens this well known Regge-trajectory-like
statement and makes an additional claim about the numerators of
conformal blocks, not only denominators: {\bf the Nekrasov expansion
provides a parametrization of the numerators by $\mu$-variables,
which are also linear functions of $\alpha$-parameters}.

Only after making this kind of presentation at level one
and for the simplest case of $c=2$ in s.2,
we turn to the standard approach: calculate conformal blocks
by the standard CFT methods in s.3,
what also justifies the results of s.2.
The necessary Ward identities, providing recursive relations
for the $3$-point functions, are taken from a separate
summary in \cite{MMMM}.
Then in s.4 we extend the check of the $U(3)$ AGT relation
to $c\neq 2$ and in s.5 to level two. At last,
in s.6 we provide the {\bf complete}
proof of the AGT relation in the very particular case of one
of the special
states in the conformal block being completely degenerate
at the first level.
The proof is possible, since there is a complete answer for the conformal
block in this case \cite{FLit}.

\section{The simplest example: two free fields with $c=2$
\label{c2exa}}

\subsection{Structure of the $W_3$ algebra}

In the case of several free fields the set of Virasoro
primaries is not exhausted by exponentials
$e^{\sqrt{2}\vec\alpha\vec\phi}$: there are many more primaries.
If one wants exponentials to remain the only primary
fields (in addition to the currents $\partial\vec\phi$),
they should be primaries of a larger algebra, extending
the Virasoro one.
The standard choice in the case of $r$ free fields
is the $W_{r+1}$ algebra.

Generators
$W^{(k)}(z) = \omega^{i_1\ldots i_k}\p\phi_{i_1}\ldots\p\phi_{i_k}(z)$
and tensors $\omega$ are defined so that only minimal singularities
are allowed in the algebra.
From now on, we switch to the simplest example of two free fields,
$r=2$, associated with the $SU(3)$ group and $W_3$ algebra.
The $W_3$ algebra has two generators: the stress tensor
$T(z) = W^{(2)}(z) = \frac{1}{2}(\partial\vec\phi)^2(z)$
and $W(z) = W^{(3)}(z)$. The stress tensor $T(z)$ is invariant under
$SO(2)$ rotations of $\phi$, and this rotation freedom
should be fixed in order to define $W(z)$ unambiguously.
We require that it is symmetric under $\phi_2 \rightarrow -\phi_2$
and, therefore, antisymmetric under $\phi_1\rightarrow -\phi_1$.
This means that $W = (\p\phi_1)^3 + h \p\phi_1(\p\phi_2)^2$
with the single undefined parameter $h$.
The operator product expansion is
\be
T(z)W(0) = \frac{3+h}{z_4}\p\phi_1(0) + \ldots
\ee
and $h$ is fixed by the requirement of absence the most singular term with $z^{-4}$
(which is equivalent to the requiring the $W$-field to be Virasoro primary).
This defines $h$ to be $h=-3$ and
\be
W(z) = W^{(3)}(z) \sim (\p\phi_1)^3 -3\p\phi_1(\p\phi_2)^2
= \p\phi_1\Big((\p\phi_1)^2 - 3(\p\phi_2)^2\Big)
\ee
(the normalization coefficient is actually $2^{-3/2}$, see
the Appendix).
Accordingly,
\be
L_0 e^{\sqrt{2}(\alpha\phi_1 + \beta\phi_2)}
= \Delta_{\alpha,\beta}e^{\sqrt{2}(\alpha\phi_1 + \beta\phi_2)},\nn \\
W_0 e^{\sqrt{2}(\alpha\phi_1 + \beta\phi_2)}
= w_{\alpha,\beta}e^{\sqrt{2}(\alpha\phi_1 + \beta\phi_2)}
\ee
with
\be
\Delta_{\alpha,\beta} = \alpha^2+\beta^2,\nn \\
w_{\alpha,\beta} = \alpha(\alpha^2-3\beta^2)
\ee
Crucial for the AGT relation in the $SU(3)$ case is
the fact, that
\be
\boxed{
\Delta_{\alpha,\beta}^3 - w_{\alpha,\beta}^2 =
\Big(\alpha^2+\beta^2\Big)^3 - \Big(\alpha(\alpha^2-3\beta^2)\Big)^2
= \Big(\beta(\beta^2-3\alpha^2)\Big)^2 = w_{\beta,\alpha}^2
}
\label{Deltaw}
\ee
The r.h.s. has nothing to do with the $W_3$ algebra, but
this relation demonstrates that a certain linear combination of
$\Delta$ and $w$, that is, the one at the l.h.s. is a full square.

\subsection{$4$-point conformal block for the $W_3$ algebra
\label{WAL3}}

As reviewed in \cite{MMMagt} and (in far more detail) in \cite{MMMM},
the conformal block has the structure
\be
{B}^{\vec\alpha}_{\vec\alpha_1,\vec\alpha_2,\vec\alpha_3,\vec\alpha_4}(x)
=\sum_{|{\cal Y}=|{\cal Y}'|} x^{|{\cal Y}|}
{\cal B}^{\vec\alpha}_{\vec\alpha_1,\vec\alpha_2,\vec\alpha_3,\vec\alpha_4}
({\cal {\cal Y}},{\cal Y}')
=\sum_{|{\cal Y}|=|{\cal Y}'|} x^{|{\cal Y}|}
\bar{\Gamma}_{\vec\alpha_1,\vec\alpha_2;\vec\alpha}({\cal Y})
Q_{\vec\alpha}^{-1}({\cal Y},{\cal Y}')
\Gamma_{\vec\alpha,\vec\alpha_3,\vec\alpha_4}({\cal Y}')
\label{cbexp}
\ee
Here ${\cal Y}$ labels the elements of the Verma module for the $W_3$ algebra,
i.e. a  generalization to $W_3$ of what was the Young diagram
for the Virasoro algebra, $\bar\Gamma({\cal Y})$ and
$\Gamma({\cal Y})$ are the two types of the corresponding
$3$-point functions
(they are essentially the same in the Virasoro case, but interrelations
are more sophisticated in the $W$-sector \cite{MMMM})
and $Q({\cal Y},{\cal Y}')$ is the Shapovalov form:
a table made from the scalar products of different elements
in the Verma module, which,
as usual, does not mix different levels $|{\cal Y}|$.

At level one, $|{\cal Y}|=1$ there are exactly two states in the Verma module,
$L_{-1}V_{\vec\alpha}$ and $W_{-1}V_{\vec\alpha}$
and the corresponding block of the Shapovalov form is
\be
\begin{array}{|c||c|c|}
\hline
&&\\
Q_{\vec\alpha}({\cal Y},{\cal Y}')
& L_{-1}V_{\vec\alpha} & W_{-1}V_{\vec\alpha} \\
&&\\
\hline\hline
&&\\
L_{-1}V_{\vec\alpha} & 2\Delta_{\vec\alpha} & 3w_{\vec\alpha}  \\
&&\\
\hline
&&\\
W_{-1}V_{\vec\alpha} & 3w_{\vec\alpha} & q\Delta^2_{\vec\alpha} \\
&&\\
\hline
\end{array}
\label{SF}
\ee
It is straightforwardly evaluated from the commutation relations
of the algebra.
However, at level one of the $W_3$-algebra the entries of
this simple matrix are defined on dimensional grounds up
to the coefficients. Coefficients $2$ and $3$ are dictated by
the elementary part of $W_3$ commutators,
\be
\left[L_n,L_m\right] = (n-m)L_{m+n} + \frac{c}{12}n(n^2-1), \nn \\
\left[L_n,W_m\right] = (2n-m)W_{m+n}
\label{LWcom}
\ee
while $q$ follows from the sophisticated one
for $\left[W_m,W_n\right]$,
which also depends on the normalization of $W(z)$.
This remaining commutation relation will appear only in
(\ref{WWcom}) far below.
Instead of using it now, one can just {\it guess} that $q=9/2$
by looking at (\ref{Deltaw}):
then the determinant $2q\Delta^3-9w^2$
of the Shapovalov form (Kac determinant),
which stands in the denominator of conformal block at level one,
factorizes nicely: moreover, it is a full square.

For this value of $q=9/2$
\be
{B}^{\vec\alpha}_{\vec\alpha_1,\vec\alpha_2,\vec\alpha_3,\vec\alpha_4}(x)
= 1 + \frac{x K_1}{\Delta_{\vec\alpha}^3 - w_{\vec\alpha}^2} + O(x^2)
\ee
or
\be
B_1 = \frac{K_1}{\Delta_{\vec\alpha}^3  - w_{\vec\alpha}^2}
 \ \stackrel{(\ref{Deltaw})}{=}\
\frac{K_1}{v^2}
\label{B1c0}
\ee
where $v^2$ is the r.h.s. of (\ref{Deltaw}) and
all dependencies on $\vec\alpha_1,\ldots,\alpha_4$ are contained
in the numerator $K_1$.

\subsection{Nekrasov's formulas}

Nekrasov's partition function has the form
\be
{\cal Z} = \sum_{Y_1,\ldots,Y_N} x^{|Y_1|+\ldots + |Y_N|}
{\cal Z}(Y_1,\ldots,Y_N)
\ee
where the sum is over $N$ sets of the {\it ordinary} Young diagrams.
The level one contribution to ${\cal Z}$ is simply
\be
{\cal Z} = 1 + x \sum_{i=1}^N
{\cal Z}(\emptyset,\ldots,\Box,\ldots\emptyset) + O(x^2)
\ee
where the only non-empty diagram $Y=\Box$ stands at the $i$-th place.
Explicitly, for the conformal invariant case $N_f= 2N$ one has
\be
{Z}_1 = \sum_{i=1}^N
{\cal Z}(\emptyset,\ldots,\Box,\ldots\emptyset)
= -\frac{1}{\epsilon_1\epsilon_2}
\sum_{i=1}^N \frac{P(a_i)}
{\prod_{j\neq i}^N(a_i-a_j)(a_i-a_j+\epsilon)}
\label{Z1gen}
\ee
where $P(a_i)=\prod_{f=1}^{2N} (a_i+\mu_f)$.
In our simple example $\epsilon=0$ and ${Z}_1$ has square of the
Van-der-Monde determinant
$\Delta(\vec a) = \prod_{i<j}(a_i-a_j)$
in the denominator,
\be
Z_1(\epsilon=0) =
\frac{M_1}{\Delta(\vec a)^2}
\label{Z1c0}
\ee
We preserve the standard notation $\Delta$ for both dimensions
and the Van-der-Monde determinants, hopefully this will not cause
a confusion. In the dimension, the argument $\vec\alpha$ is a subscript,
while in the determinant $\vec a$ is an argument in brackets.
The vector $\vec a$ is actually $r=N-1$-dimensional, like $\vec\alpha$,
since its $N$ components $a_i$
(eigenvalues of the gauge-field vev matrix)
are constrained by the zero trace
condition
\be\sum_{i=1}^N a_i = 0
\label{trless}
\ee

\subsection{AGT relation
\label{agt1}}

The AGT relation \cite{AGT,Wyl,MMMagt}
is the statement that for some linear relation
between the whole set of $\vec\alpha$'s and the whole set of
$\vec a$ and $\mu_f$, the conformal blocks ${\cal B}$ for
the $W_N$ algebra and the Nekrasov
partition function ${\cal Z}$ for $U(N)$ group coincide
provided $c = (N-1)\left(1+N(N+1)\frac{\epsilon^2}
{\epsilon_1\epsilon_2}\right)$.
We start from the simplest version of this
relation for $N=3$: for the
4-point conformal block and for $c=2$, i.e. $\epsilon=0$.

Comparing (\ref{B1c0}) and (\ref{Z1c0}) one observes that
in this case the AGT relation requires that
\be
v \sim \Delta(\vec a)
\label{vVDMa}
\ee
which dictates the relation between $\vec a$ and $\vec\alpha$
of the intermediate state,
and
\be
K_1 \sim M_1
\label{KM}
\ee
which would define the relation between $\mu_f$
and $\vec\alpha_1,\ldots,\vec\alpha_4$ for the four
external states.

\subsubsection{Denominator}

Let us begin with (\ref{vVDMa}).
Coming back to the parametrization $\vec\alpha = (\alpha,\beta)$,
one obtains
\be
v \ \stackrel{(\ref{Deltaw})}{=}\ \beta(\beta^2-3\alpha^2)
= 4\beta \cdot \frac{\beta + \alpha\sqrt{3}}{2}
\cdot \frac{\beta - \alpha\sqrt{3}}{2} \sim
(a_1-a_2)(a_2-a_3)(a_3-a_1)
\label{va}
\ee
which, together with (\ref{trless}), implies a linear relation
of the form
\be
a_1 = \kappa (\alpha-\beta\sqrt{3}), \nn \\
a_2 = \kappa (\alpha+\beta\sqrt{3}), \nn \\
a_3 = -2\kappa\alpha\phantom{\alpha+\beta\ \ }
\label{avsalphabeta}
\ee
It can be, of course, written in terms of the root and
weight vectors of $SU(3)$.

\subsubsection{Numerator.
The case of $\ \vec\alpha_1=\ldots=\vec\alpha_4=0$}

Let us now proceed to (\ref{KM}).
We begin with putting all the four "external" dimensions equal to zero,
$\alpha_1=\beta_1=\ldots=\alpha_4=\beta_4=0$.
Then (\ref{KM}) follows from
\be
B^{(1)} = \frac{\Delta(p\Delta^3 + sw^2)}{\Delta^3-w^2} =
-\frac{1}{\epsilon_1\epsilon_2}\left(\frac{P_6(a_1)}{a_{12}^2a_{23}^2} +
\frac{P_6(a_2)}{a_{21}^2a_{23}^2} +
\frac{P_6(a_3)}{a_{31}^2a_{32}^2}\right) = Z^{(1)}
\label{ZB1}
\ee
where only the two parameters $p$ and $s$ at the l.h.s.
are not defined on dimensional grounds and
$P_6(a)=a^6 + \sum_{k=1}^6\sigma_k a^{6-k}$
at the r.h.s. is a polynomial of degree $6$, depending on
choice of the $6$ parameters $\mu_f$ through symmetric
polynomials
\be
\sigma_k = \sum_{f_1<\ldots<f_k}\mu_{f_1}\ldots\mu_{f_k}
\label{sigmu}
\ee
As usual, $a_{ij} \equiv a_i-a_j$.
Like the $SU(2)$ case, \cite{MMMagt}
the parameter $-\epsilon_1\epsilon_2$ should be absorbed into rescaling
of $\Delta$ and $w$.
Now one substitutes $w^2 = \Delta^3-v^2$ with
\be
\Delta = -\frac{(\alpha^2+\beta^2)}{\epsilon_1\epsilon_2}
\ \stackrel{(\ref{avsalphabeta})}{=}\
-\frac{1}{6\kappa^2\epsilon_1\epsilon_2}(a_1^2+a_2^2+a_3^2)
\ee
and
\be
v^2 \ \stackrel{(\ref{Deltaw})}{=}\
-\frac{\Big(\beta(\beta^2-3\alpha^2)\Big)^2}{(\epsilon_1\epsilon_2)^3}
 \ \stackrel{(\ref{avsalphabeta})}{=}\
-\frac{1}{108\kappa^6(\epsilon_1\epsilon_2)^3}
(a_1-a_2)^2(a_2-a_3)^2(a_3-a_1)^2
\ee
and obtains for (\ref{KM})
\be
(a_1^2+a_2^2+a_3^2)\Big( \tilde p\,(a_1^2+a_2^2+a_3^2)^3 -
\tilde s a_{12}^2a_{23}^2a_{31}^2 \Big)
=
a_{23}^2P_6(a_1) +
a_{13}^2P_6(a_2) +
a_{23}^2P_6(a_3)
\ee
where $\tilde p = \frac{p+s}{12\kappa^2}$,
$\tilde s = \frac{s}{6\kappa^2}$, and
one should also impose condition (\ref{trless}),
i.e. substitute $a_3=-a_1-a_2$.
When all the external dimensions are vanishing
(we still keep $\epsilon=0$),
the $\mu$-parameters at the r.h.s. are vanishing as well:
\be
\vec\alpha_1=\ldots=\vec\alpha_4=\epsilon=0
\ \Longrightarrow\ \mu_1=\ldots=\mu_6=0
\ee
Then the difference between the l.h.s. and the r.h.s. is
a polynomial of degree $8$ in $a_1$ and $a_2$, equal to
\be
2a_1^2a_2^2\Big((80\tilde p - 13\tilde s - 4)(a_1^4+a_2^4)
+ (128\tilde p + 17\tilde s +2)a_1a_2(a_1^2+a_2^2)
+ (152\tilde p + 32\tilde s + 5)a_1^2a_2^2\Big)
\ee
For the AGT relation to be true this expression should vanish
identically in $a_1$ and $a_2$, i.e. the three coefficients
should be made vanishing by adjusting just the two parameters
$\tilde p$ and $\tilde s$.
Surprisingly or not, the three equations are indeed consistent,
and their common solution is
$\tilde p = 1/72$, $\tilde s = -2/9$.
This corresponds to $s/p=-8/9$, i.e. one obtains (\ref{ZB1})
in the form of the algebraic identity
\be
\boxed{
B^{(1)} = \frac{3\kappa^2}{2}\cdot
\frac{\Delta\left(\Delta^3-\frac{8}{9}w^2\right)}
{\Delta^3-w^2} \ \stackrel{(\ref{avsalphabeta})}{=}\
-\frac{1}{\epsilon_1\epsilon_2}\left(\frac{a_1^6}{a_{12}^2a_{23}^2} +
\frac{a_2^6}{a_{21}^2a_{23}^2} +
\frac{a_3^6}{a_{31}^2a_{32}^2}\right) = Z^{(1)}
}
\label{B1c2}
\ee
The conformal block at the l.h.s. can be of course calculated from
representation theory of the $W_3$ algebra, as we shall see
in s.\ref{CFT} below.

\subsubsection{Numerator. The case of arbitrary $\vec\alpha_2$
and $\vec\alpha_4$ with no ambiguity in the $W$-conformal block
}

If one switches on dimensions, the l.h.s. of (\ref{ZB1}) gets
more complicated and additional contributions should be compensated
by adjustment of six $\mu$'s at the r.h.s.
According to (\ref{cbexp}) and (\ref{SF}) with $q=9/2$, the
conformal block at the l.h.s. is now given by
\be
2(\Delta_{\vec\alpha}^3-w_{\vec\alpha}^2)B^{(1)} = \Delta^2
\bar\Gamma_{\vec\alpha_1,\vec\alpha_2;\vec\alpha}(L_{-1})
\Gamma_{\vec\alpha_3,\vec\alpha_4;\vec\alpha}(L_{-1})-
\nn
\ee
\be
- \frac{2w_{\vec\alpha}}{3}\Big(
\bar\Gamma_{\vec\alpha_1,\vec\alpha_2;\vec\alpha}(L_{-1})
\Gamma_{\vec\alpha_3,\vec\alpha_4;\vec\alpha}(W_{-1}) +
\bar\Gamma_{\vec\alpha_1,\vec\alpha_2;\vec\alpha}(W_{-1})
\Gamma_{\vec\alpha_3,\vec\alpha_4;\vec\alpha}(L_{-1})\Big)
+ \frac{4\Delta_{\vec\alpha}}{9}
\bar\Gamma_{\vec\alpha_1,\vec\alpha_2;\vec\alpha}(W_{-1})
\Gamma_{\vec\alpha_3,\vec\alpha_4;\vec\alpha}(W_{-1})
\label{cbexp1}
\ee
where
\be
\bar\Gamma_{\vec\alpha_1,\vec\alpha_2;\vec\alpha}(L_{-1}) =
\Delta_{\vec\alpha}+\Delta_{\vec\alpha_1}-\Delta_{\vec\alpha_2}
\ee
and $\Gamma_{\vec\alpha_3,\vec\alpha_4;\vec\alpha}(L_{-1})$ is given by exactly
the same formula,
while $\Gamma (W_{-1})$ and $\bar\Gamma (W_{-1})$
require additional restrictions to be uniquely determined
as was explained in the Introduction.

We begin with the case of $\vec\alpha_1=\vec\alpha_3=\vec 0$,
when the ambiguity is known not to show up.
Then
\be
\Gamma_{\vec 0,\vec\alpha_4;\vec\alpha}(W_{-1}) =
\xi w_{\vec\alpha} + \eta w_{\vec\alpha_2},\ \ \ \ \ \ \
\bar\Gamma_{\vec 0,\vec\alpha_2;\vec\alpha}(W_{-1}) =
\bar\xi w_{\vec\alpha} + \bar\eta w_{\vec\alpha_2}
\label{omegaw2}
\ee
with some coefficients $\xi$, $\bar\xi$, $\eta$ and $\bar\eta$.
In order to find $\xi$, one can also put $\vec\alpha_2=\vec\alpha_4=0$,
then (\ref{cbexp1}) turns into
\be
B^{(1)} = \frac{\Delta}{2}\cdot
\frac{\Delta^3+w^2\left(\left(\frac{2\xi}{3}-1\right)
\left(\frac{2\bar\xi}{3}-1\right)
- 1\right)}
{\Delta^3-w^2}
\label{B1c2CFT}
\ee
Comparison with (\ref{B1c2}) implies that $\boxed{3\kappa^2=1}$
and $\boxed{\left(\frac{2\xi}{3}-1\right)
\left(\frac{2\bar\xi}{3}-1\right)=1/9}$.

Thus, we see that the AGT relation together with the
basic definition (\ref{cbexp}) can only partly fix
the coefficients in formulas for the conformal block at level one.
Therefore, we glance at s.3.2 and fix the values $\xi=\bar\xi$.
Then, $\xi=1$ or $2$. The formulas of s.3.2 fix $\xi=1$.
Then, under switching on $\vec\alpha_2$ and $\vec\alpha_4$, (\ref{B1c2CFT})
becomes
\be
B^{(1)} = \frac{9\Delta^2(\Delta-\Delta_2)(\Delta-\Delta_4)
-6w(w+\eta w_2)(\Delta-\Delta_4)
-6w(w+\eta w_4)(\Delta-\Delta_2)
+ 4\Delta(w+\eta w_2)(w+\eta w_4)}{18(\Delta^3-w^2)}
\nn
\ee
which should be now compared with the r.h.s. of (\ref{ZB1}).
The resulting equations are identically satisfied for any $a$, provided
\be
\sigma_1 = 0, \nn \\
\sigma_2 =  -\Big(\alpha_2^2+\beta_2^2+\alpha_4^2+\beta_4^2\Big),\nn\\
\sigma_3= -\frac{2}{3\sqrt{3}}
\Big(\bar\eta\alpha_2(\alpha_2^2-3\beta_2^2)
+\eta\alpha_4(\alpha_4^2-3\beta_4)^3\Big),\nn\\
\sigma_4 =
(\alpha_2^2+\beta_2^2)(\alpha_4^2+\beta_4^2),\nn\\
\sigma_5 = \frac{2}{3\sqrt{3}}\Big(\eta
(\alpha_2^2+\beta_2^2)\alpha_4(\alpha_4^2-3\beta_4^2)
+\bar\eta(\alpha_4^2+\beta_4^2)\alpha_2(\alpha_2^2-3\beta_2^2)\Big),\nn\\
\sigma_6 = \frac{4\eta\bar\eta}{27}\ \alpha_2(\alpha_2^2-3\beta_2^2)
\alpha_4(\alpha_4^2-3\beta_4^2)
\label{sigxi1}
\ee
However, we still need to decompose these $\sigma$'s in $\mu$
according to (\ref{sigmu}).
The AGT relation requires $\mu$'s to be linear functions of $\vec\alpha$'s,
and it is clear from the simple formula for $\sigma_6$ that
this is indeed the case, with
\be
\mu_1=-\bar\eta\frac{2}{\sqrt{3}}\alpha_2, \nn\\
\mu_2=\bar\eta\frac{\alpha_2+\beta_2\sqrt{3}}{\sqrt{3}}, \nn\\
\mu_3=\bar\eta\frac{\alpha_2-\beta_2\sqrt{3}}{\sqrt{3}}, \nn\\
\mu_4=-\eta\frac{2}{\sqrt{3}}\alpha_4, \nn\\
\mu_5=\eta\frac{\alpha_4+\beta_4\sqrt{3}}{\sqrt{3}}, \nn\\
\mu_6=\eta\frac{\alpha_4-\beta_4\sqrt{3}}{\sqrt{3}}
\label{mu1c2}
\ee
and $\boxed{\eta,\ \bar\eta=\pm 1}$.
This constraint is not a surprise
since in (\ref{sigxi1}) $\eta$ and $\bar\eta$ appear rather
irregularly.
This solution also reproduces all other $\sigma$'s
and it is unique up to the sign choice for $\eta$, $\bar\eta$
and up to $6!$ permutations of $\mu$'s.

\subsubsection{The case of $\xi=2$}

Let us now see what happens if one chooses $\xi=\bar\xi=2$ in (\ref{omegaw2}).
It turns out that this time the overdefined system
of equations for six $\sigma$'s have {\it no} solution.
Solution appears if one takes into account an extra factor
\be
Z^{U(1)} = (1-x)^{-\nu}
\ee
in the AGT relation,
\be
{\cal B} = Z^{U(1)}Z^{SU(N)}
\ee
At level one introduction of $\nu$ implies that
(\ref{ZB1}) is deformed into
\be
B^{(1)} = Z^{(1)} + \nu
\label{ZB1nu}
\ee
If $\nu\neq 0$ is allowed, then $\xi=2$ also provides
a solution to (\ref{ZB1nu}), actually with
the same parameters $\mu$ as in (\ref{mu1c2}) and with
\be\label{nuintch}
\nu = \alpha_2^2+\alpha_4^2+\beta_2^2+\beta_4^2=\Delta_2+\Delta_4
\ee
The origin for this other solution is that it corresponds to the other choice
of the special states, 2 and 4. This is why this value of $\xi$ is not consistent
with the table (\ref{vertices}) where the special states are chosen to be 1 and 3.

\subsubsection{The $U(1)$ factor $\nu$ and projective
transformations}

Thus we observe the same phenomenon as in the $SU(2)$
case in \cite{MMMagt}: if one makes the simultaneous interchange of
external lines $\vec\alpha_1 \leftrightarrow \vec\alpha_2$
and $\vec\alpha_3 \leftrightarrow \vec\alpha_4$
a simple $\nu$ (zero in the particular case of
$\vec\alpha_1=\vec\alpha_3=0$ which we are now considering)
turns into a sophisticated one.
In principle this is nothing but a result of modular
transformation of the conformal block. Indeed, the generic four-point
function is \cite{BPZ,ZZ}
\be\label{4V}
\langle V_{\Delta_1}(z_1)V_{\Delta_2}(z_2)V_{\Delta_3}(z_3)V_{\Delta_4}(z_4)
\rangle =\prod_{i<j}^4 (z_i-z_j)^{d_{ij}}(\bar z_i-\bar z_j)^{\bar d_{ij}}G(x,\bar x)
\ee
where $d_{34}=d_{13}=0$, $d_{14}=-2\Delta_2$, $d_{24}=\Delta_1+\Delta_3-
\Delta_2-\Delta_4$, $d_{34}=\Delta_1+\Delta_2-\Delta_3-\Delta_4$, $d_{23}=
\Delta_4-\Delta_2-\Delta_1-\Delta_3$, $x={(z_1-z_2)(z_3-z_4)\over
(z_3-z_2)(z_1-z_4)}$ and similarly for the complex conjugated part.
$G(x,\bar x)$ is a bilinear combination of the conformal blocks
$B(\{\Delta_i\},\Delta,,c;x)$ and
$B(\{\bar\Delta_i\},\bar\Delta, c;\bar x)$.
Due to the projective invariance, one can choose three of these four
points arbitrarily. If choosing $z_4=\infty$, $z_3=1$ and $z_2=0$,
one obtains $z_1=x$ and the 4-point correlator (\ref{4V}) becomes
\be
(1-x)^{d_{12}}\bar x^{\bar d_{13}}(1-\bar x)^{\bar d_{12}}x^{d_{13}}G(x,\bar x)
\ee
The simultaneous interchange of
external lines $\vec\alpha_1 \leftrightarrow \vec\alpha_2$
and $\vec\alpha_3 \leftrightarrow \vec\alpha_4$ in formula (\ref{4V})
does not change $x$ and $\bar x$
and leads to the factor of $(1-x)^{d_{24}}$ instead of $(1-x)^{d_{13}}$ ($d_{13}=0$)
which is exactly $\nu=-d_{24}$ in (\ref{nuintch})
(since $\Delta_1=\Delta_3=0$ there).

\subsubsection{General case, all $\vec\alpha\neq 0$,
$\vec\alpha_1$ and $\vec\alpha_3$ {\it special}}

We now come to the general case of all external $\vec\alpha$ switched on.
In this case, one needs some ansatz for the $3$-point functions $\Gamma (W_{-1})$,
$\bar\Gamma (W_{-1})$. However, the number of parameters is large enough
and the adjusting procedure fails to be very
effective to define the coefficients. As we know from the
$SU(2)$ case, restrictions get very strong when one includes
the AGT relation at higher levels \cite{MMMagt}, see s.5.

Therefore, we return to a more straightforward approach and
just take the true value of the $3$-point function from
the conformal field theory analysis, see \cite{MMMM} and s.\ref{3pf} below.
Thus, we take the vertices from s.3.2 and
fix the special states to have $\beta_{1,3} = r_{1,3}\alpha_{1,3}$
with some fixed $r_{1,3}$.

Then AGT relation is satisfied, provided there is a certain
linear relation between the sextuplets of parameters
$\alpha_1,\vec\alpha_2,\alpha_3,\vec\alpha_4$
and $\mu_1,\ldots,\mu_6$.
One of the possible solutions is
\be
\mu_1=\frac{2}{\sqrt{3}}(\alpha_2-\alpha_1), \nn\\
\mu_2=-\frac{\alpha_2+2\alpha_1+\beta_2\sqrt{3}}{\sqrt{3}}, \nn\\
\mu_3=-\frac{\alpha_2+2\alpha_1-\beta_2\sqrt{3}}{\sqrt{3}}, \nn\\
\mu_4=-\frac{2}{\sqrt{3}}(\alpha_4+\alpha_3), \nn\\
\mu_5=\frac{\alpha_4+2\alpha_3+\beta_4\sqrt{3}}{\sqrt{3}}, \nn\\
\mu_6=\frac{\alpha_4+2\alpha_3-\beta_4\sqrt{3}}{\sqrt{3}}
\label{mu1c2gen}
\ee
In this solution, the $U(1)$-factor shows up in the level one
AGT relation (\ref{ZB1nu}) with $\nu = -4\alpha_1\alpha_3$, while
\be\label{r}
r_{1}=-\sqrt{3},\ \ \ \ \  r_3=\sqrt{3}
\ee

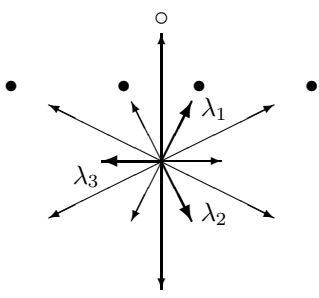
\begin{figure}
\unitlength 1mm 
\linethickness{0.4pt}
\ifx\plotpoint\undefined\newsavebox{\plotpoint}\fi 
\begin{picture}(43.108,24.13)(-90,-15)
\put(0,0){\vector(0,1){17}}
\put(0,0){\vector(0,-1){17}}
\put(0,0){\vector(2,1){15}}
\put(0,0){\vector(2,-1){15}}
\put(0,0){\vector(-2,1){15}}
\put(0,0){\vector(-2,-1){15}}
\put(0,0){\vector(1,0){8}}
\put(0,0){\vector(-1,2){4}}
\put(0,0){\vector(-1,-2){4}}
\put(20,10){\circle*{1.3}}
\put(-20,10){\circle*{1.3}}
\put(5,10){\circle*{1.3}}
\put(-5,10){\circle*{1.3}}
\put(0,19){\circle{1.3}}
\put(7,7){\makebox(0,0)[cc]{$\lambda_1$}}
\put(7,-7){\makebox(0,0)[cc]{$\lambda_2$}}
\put(-10,-2){\makebox(0,0)[cc]{$\lambda_3$}}
\thicklines
\put(0,0){\vector(-1,0){8}}
\put(0,0){\vector(1,2){4}}
\put(0,0){\vector(1,-2){4}}
\end{picture}
\caption{\footnotesize{
Roots and minimal weights of $sl(3)$. The roots have the length
$\sqrt{2}$, the minimal weights have $\sqrt{2/3}$.
All angles are integer multiples of $30^\circ$.
The two simple roots and
the corresponding two fundamental weights
are marked by black circles.
The third positive root, marked with a white circle is
also $\vec\rho$, the half-sum of all (three) positive roots
or the sum of (two) fundamental weights.
The vectors $\vec\lambda_{1,2,3}$ are shown by thick lines,
they form a Weyl-invariant triple of minimal weights.
}}
\label{picroots}
\end{figure}
Let us introduce the three vectors, see Fig.\ref{picroots}
\be\label{fundvec}
\vec\lambda_1 = \frac{1}{\sqrt{6}}\Big(1,\sqrt{3}\Big),
\ \ \ \ \
\vec\lambda_2 = \frac{1}{\sqrt{6}}\Big(1,-\sqrt{3}\Big),
\ \ \ \ \
\vec\lambda_3= \sqrt{\frac{2}{3}}\,\Big(-1,0\Big)
\ee
They are three (out of six) minimal vectors in the weight lattice
of $sl(3)$.
Then $a_i$ in (\ref{avsalphabeta}) with $\kappa=1/\sqrt{3}$
and $\mu_f$ in (\ref{mu1c2gen})
can be represented as scalar products,
\be
a_{1,2,3} = \sqrt{2}\vec\lambda_{1,2,3}\vec\alpha
= (\sqrt{2}\vec\alpha)\vec\lambda_{1,2,3}
\ee
and
\be
\mu_{1,2,3}
= -(\sqrt{2}\vec\alpha_2)\vec\lambda_{1,2,3}
- \frac{2}{\sqrt{3}}\alpha_1, \nn \\
\mu_{4,5,6}
= (\sqrt{2}\vec\alpha_4)\vec\lambda_{1,2,3}
- \frac{2}{\sqrt{3}}\alpha_3
\label{mu1c2genvect}
\ee
Let us remind that the primary exponentials are
$V_{\vec\alpha} = e^{(\sqrt{2}\vec\alpha)\vec\phi}$,
this can make these formulas looking more natural.
The eigenvalues of the $W^{(3)}$ operator are
\be
w_{\vec\alpha} = -3\sqrt{6} (\vec\alpha\vec\lambda_1)
(\vec\alpha\vec\lambda_2) (\vec\alpha\vec\lambda_3)
= \alpha(\alpha^2-3\beta^2)
\ee
Note that one could choose the pair of the {\it special} states
in a way, different from (\ref{r}), that is, $r_{1,3}=0,\pm\sqrt{3}$.
We return to complete analysis in s.\ref{muvecs} below,
and the full final result is summarized in the Conclusion,
eqs.(\ref{concmu})-(\ref{concshi}).

\subsection{Intermediate summary}

This ends the explicit check of the AGT conjecture
for the 4-point conformal block of $W_3$-algebra with $c=2$
{\it at level one}: it is indeed expressed through
Nekrasov's special functions with $\vec a$ and $\mu_f$
linearly expressed through $\vec\alpha$'s.
Technical generalizations can go in four obvious directions.

First, one can check the AGT relation between the structure
constants of conformal model and "perturbative"
Nekrasov's functions.
This relation refers to the 3-point functions
of particular conformal model:
Liouville in the $SU(2)$ and conformal Toda in the more
general $SU(N)$ case, which are not yet known
in full generality.
It was also discussed in some detail in \cite{Wyl}.

The second and third, one can either
switch on $\epsilon\neq 0$ or proceed to higher levels
or do the both, and repeat the
calculations made for the $SU(2)$ case in \cite{MMMagt}.
Fourth, one can consider the $SU(N)/W_N$ examples with $N>3$.
These are straightforward exercises in representation theory.

In order proceed to a discussion of the second
and third directions in ss.\ref{cn2} and \ref{secl}
respectively, we need some additional information
from the theory of $W$-algebras, it is reminded in
\cite{MMMM} and, more briefly, in the following section \ref{CFT}.
As to the fourth direction, we discuss only one,
but very interesting subject in s.\ref{FLpro}:
it concerns a general role of the Nekrasov functions and,
more generally, matrix-model $\tau$-functions as a modern
substitute for hypergeometric functions.
As a particular application,
s.\ref{FLpro} comments on the complete proof \cite{mmNF}
of the AGT relation for the special case when the conformal blocks
are hypergeometric functions (rather than generic "hypergeometric
integrals" \cite{SheVa,MV}, to which they are supposedly related in the
Dotsenko-Fateev approach).

\section{
Some CFT considerations
\label{CFT}}

In s.\ref{c2exa} we constructed conformal blocks at level one
mostly from dimensional considerations and demonstrated
that remaining few parameters can be adjusted so that
the AGT relations are fulfilled.
Moreover, even at level one restrictions are rather strong,
they actually become exhaustive already at level two.
We did this on purpose: to demonstrate that the AGT relations
can in the future become a powerful alternative approach
to conformal field theory problems, which are pretty hard
when attacked directly.
Still, conformal blocks are conformal blocks and they can
of course be evaluated by the standard CFT methods.
In this section we briefly sketch how to evaluate the relevant
quantities: the 3-point functions $\Gamma$ and $\bar\Gamma$
at level one and the Shapovalov matrix.
The details can be found in \cite{MMMM}.
For original papers and reviews see \cite{BPZ,ZZ,fref,FLuk,BW,FLit}.

Note that we use somewhat non-standard normalizations,
adjusted to maximally simplify the $W$-algebra formulas.
The price for this is certain deviation from
conventions in some other papers.
We illustrate our conventions with the example of free field
theory in the Appendix.

\subsection{$W$-algebra and Shapovalov form}

In s.\ref{WAL3} we defined the parameter $q$ in (\ref{SF})
from the requirement that the corresponding Kac determinant
decomposes nicely with the help of (\ref{Deltaw}).
In CFT $q$ is calculated from the structure constants
of $W_3$-algebra. However, the simple ones (\ref{LWcom})
are not sufficient, one needs also the remaining, non-linear
commutation relation \cite{ZW}-\cite{FLit}:
\be
\frac{2}{9}\left[ W_n,W_m\right] =
(n-m)\left\{\frac{16}{22+5c}\Lambda_{n+m}
+ \left(\frac{(n+m+2)(n+m+3)}{15} -\frac{(n+2)(m+2)}{6}\right)
L_{n+m}\right\} + \nn \\
+ \frac{c}{3\cdot 5!}n(n^2-1)(n^2-4)\delta_{n+m,0}
\label{WWcom}
\ee
Here $\Lambda_n$ is a remnant of the $W^{(4)}$ operator.
It does not exist as an independent operator in the $W_3$ algebra
for the same reason that
$\ {\rm tr}\, J^4 = \frac{1}{2}\left({\rm tr}\, J^2\right)^2$
for a $3\times 3$ traceless matrix $J$. In fact,
\be
\Lambda_n = \sum_{k=-\infty}^\infty :L_kL_{n-k}:\ +\
\frac{x_n}{5}L_n,
\ee
with $x_{2l} = 1-l^2$,\ \ \ $x_{2l+1} = (2+l)(1-l)$.
We use the notation from \cite{FLit}, but
our normalization of the $W$-operator is different
by a factor of $3/\sqrt{2}$.
Note for $n=0$ that the term with the normal ordering
contains an item $L_0^2$ which {\it does} contribute to
the vacuum average:
\be
<V_\alpha \Lambda_0 V_\alpha> =
\Delta_\alpha^2 + \frac{1}{5}\Delta_\alpha =
\Delta_\alpha\left(\Delta_\alpha+\frac{1}{5}\right)
\ee
Thus one obtains for the "difficult" element of the Shapovalov
matrix (\ref{SF}),
(see \cite{MMMagt} where a similar calculation
for the Virasoro group is reminded in detail):
\be
<W_{-1}V_{\vec\alpha}\,|W_{-1}V_{\vec\alpha}>\ =\
<V_{\vec\alpha}\,|\, [W_1,W_{-1}] V_{\vec\alpha}>\ =
\frac{9\Delta_{\vec\alpha}}{2}\left(\frac{32}{22+5c}
\Big(\Delta_{\vec\alpha}+\frac{1}{5}\Big) - \frac{1}{5}\right)
\equiv \frac{9D_{\vec\alpha}\Delta_{\vec\alpha}}{2}
\ \ \stackrel{c=2}{\longrightarrow}\ \
\frac{9}{2}\Delta_{\vec\alpha}^2
\ee
in accordance with the {\it guess} in s.\ref{WAL3}.
The other three entries of the table (\ref{SF}) do {\it not}
depend on $c$.
Likewise, more generally,
\be
<W_{-n}V_{\vec\alpha}\,|W_{-n}V_{\vec\alpha}>\ =\
\frac{9n\Delta_{\vec\alpha}}{2}\left(\frac{32}{22+5c}
\Big(\Delta_{\vec\alpha}+\frac{1}{5}\Big) + \frac{5n^2-8}{15}\right)
= \frac{9D_{\vec\alpha}\Delta_{\vec\alpha}}{2}
= \frac{9n\Delta_{\vec\alpha}}{2}
\left(D_{\vec\alpha}+\frac{n^2-1}{3}\right)
\label{WnWn}
\ee
We introduced here a peculiar parameter
\be
D_{\vec\alpha} \equiv \frac{32}{22+5c}
\Big(\Delta_{\vec\alpha}+\frac{1}{5}\Big) - \frac{1}{5}
\label{Ddef}
\ee
which helps to get some formulas at low levels shorter
(however, $D$ is nothing like a universal effective variable
to absorb, say, the $c$-dependence; at most, it is a convenient
abbreviation for limited purposes).
Two other expressions of this kind are:
\be\label{WnW-n}
\left[W_0,W_{-1}\right] V_{\vec\alpha}
= \frac{9D}{2}L_{-1}V_{\vec\alpha},\nn\\
W_1W_{-1}V_{\vec\alpha} = \left[W_1,W_{-1}\right] V_{\vec\alpha}
= \frac{9D\Delta}{2}V_{\vec\alpha}
\ee
where $V_{\vec\alpha}$ is arbitrary primary.

\subsection{$3$-point functions at level one
\label{3pf}}

Three point functions of interest for us
are evaluated simply by moving integration contours.
In (\ref{cbexp}) we need the two {\it types} of such functions,
$\Gamma$ and $\bar\Gamma$, which are different in the $W$-sector,
see \cite{MMMM}.

We begin with the $\Gamma (L_{1})$, i.e. with the {\it stress-tensor}
insertion into a {\it correlator} of three primary fields
(we write it for the operators $V_3$ and $V_4$, rather than $V_1$ and
$V_2$, since this is the form in which we actually need it
in (\ref{cbexp}),
also location of the Virasoro operator is underlined to make
the formula readable):
\be
\langle \underline{L_nV_\alpha(0)}
\ V_{\alpha_3}(1)\ V_{\alpha_4}(\infty)\rangle\ =
\oint_0 x^{n+1}dx\
\langle T(x) V_\alpha(0)\ V_{\alpha_3}(1)\ V_{\alpha_4}(\infty)\rangle\
= \nn \\ = -\left(\oint_1 x^{n+1}dx\ + \oint_\infty x^{n+1}dx\right)
\langle T(x) V_\alpha(0)\ V_{\alpha_3}(1)\ V_{\alpha_4}(\infty)\rangle\
= \nn \\ = -\sum_k \oint_1 \frac{x^{n+1}dx}{(x-1)^{k+2}}
\langle V_\alpha(0)\
\underline{L_k V_{\alpha_3}(1)}\ V_{\alpha_4}(\infty)\rangle\
+ \sum_k \oint_\infty x^{k-2} x^{n+1}dx
\langle V_\alpha(0)\  V_{\alpha_3}(1)\
\underline{L_kV_{\alpha_4}(\infty)}\rangle\
= \nn \\
= -(n+1)\Delta_3
\langle V_\alpha(0)\ V_{\alpha_3}(1)\ V_{\alpha_4}(\infty)\rangle\
- \langle V_\alpha(0)\ \underline{L_{-1}V_{\alpha_3}(1)}
\ V_{\alpha_4}(\infty)\rangle\
+ \langle V_\alpha(0)\ V_{\alpha_3}(1)\
\underline{L_{-n}V_{\alpha_4}(\infty)}\rangle\
\label{Lncont}
\ee
where we used the fact that primaries are annihilated by $L_n$
with $n>0$ and the action of $L_0$ produces the dimension:
because of this, the first sum in the third line is reduced to two terms. Only
one term contributes to the second sum even without these conditions.
Now one can substitute $n=-1$, $n=0$ and $n=1$ to get
(in an obvious abbreviated notation)
\be
\langle L_{-1} V_\alpha(0) \rangle =
- \langle L_{-1} V_{\alpha_3}(1) \rangle, \nn \\
\Delta = -\Delta_3 - \langle L_{-1} V_{\alpha_3}(1) \rangle + \Delta_4,\nn\\
0 = -2\Delta_3 - \langle L_{-1} V_{\alpha_3}(1) \rangle
+  \langle L_{-1} V_{\alpha_4}(\infty) \rangle
\ee
which can be resolved for any of the three positions of the
$L_{-1}$ operator:
\be
\langle \underline{L_{-1}V_\alpha(0)}
\ V_{\alpha_3}(1)\ V_{\alpha_4}(\infty)\rangle\ =
\Big(\Delta+\Delta_3-\Delta_4\Big)
\langle V_\alpha(0)\ V_{\alpha_3}(1)\ V_{\alpha_4}(\infty)\rangle, \nn \\
\boxed{
\langle V_\alpha(0)\ \underline{L_{-1}V_{\alpha_3}(1)}
\ V_{\alpha_4}(\infty)\rangle\ =
\Big(-\Delta-\Delta_3+\Delta_4\Big)
\langle V_\alpha(0)\ V_{\alpha_3}(1)\ V_{\alpha_4}(\infty)\rangle,
} \nn \\
\langle V_\alpha(0)\ V_{\alpha_3}(1)\
\underline{L_{-1}V_{\alpha_4}(\infty)}\rangle\ =
\Big(-\Delta + \Delta_3 + \Delta_4\Big)
\langle V_\alpha(0)\ V_{\alpha_3}(1)\ V_{\alpha_4}(\infty)\rangle
\label{L-1rels}
\ee
It is the first position that we denote through
$\Gamma_{\alpha_3,\alpha_4;\alpha}(L_{-1})$
in the main text, and it is the second (boxed) one that
we use when constructing $\Gamma_{\alpha_3,\vec\alpha_4;\vec\alpha}(W_{-1})$
for the special state $\vec\alpha_3$.

Similarly, the $\Gamma (W_{-1})$-function results from the
study of $W$-operator insertions:
\be
\langle \underline{W_nV_\alpha(0)}
\ V_{\alpha_3}(1)\ V_{\alpha_4}(\infty)\rangle\ =
\oint_0 x^{n+2}dx\
\langle W(x) V_\alpha(0)\ V_{\alpha_3}(1)\ V_{\alpha_4}(\infty)\rangle\
= \nn \\ = -\left(\oint_1 x^{n+2}dx\ + \oint_\infty x^{n+2}dx\right)
\langle W(x) V_\alpha(0)\ V_{\alpha_3}(1)\ V_{\alpha_4}(\infty)\rangle\
= \nn \\ = -\sum_k \oint_1 \frac{x^{n+2}dx}{(x-1)^{k+3}}
\langle V_\alpha(0)\
\underline{W_k V_{\alpha_3}(1)}\ V_{\alpha_4}(\infty)\rangle\
\boxed{-}\ \sum_k \oint_\infty x^{k-3} x^{n+2}dx
\langle V_\alpha(0)\  V_{\alpha_3}(1)\
\underline{W_kV_{\alpha_4}(\infty)}\rangle\
= \nn \\
=-\frac{(n+1)(n+2)w_3}{2}
\langle V_\alpha(0)\ V_{\alpha_3}(1)\ V_{\alpha_4}(\infty)\rangle\
- (n+2)
\langle V_\alpha(0)\
\underline{W_{-1}V_{\alpha_3}(1)}\ V_{\alpha_4}(\infty)\rangle\
- \nn \\
- \langle V_\alpha(0)\
\underline{W_{-2}V_{\alpha_3}(1)}\ V_{\alpha_4}(\infty)\rangle\
- \langle V_\alpha(0)\ V_{\alpha_3}(1)\
\underline{W_{-n}V_{\alpha_4}(\infty)}\rangle\
\label{Wncont}
\ee
Note that the sign in the box is {\it different} from the
Virasoro case, because the odd-spin $3$-differential $W(z)$
is transformed
differently from the even-spin $2$-differential $T(z)$
under the change
$z\rightarrow 1/z$, $dz/z\rightarrow -dz/z$.
Applying the same trick, i.e. putting $n=-1,0,1$ one gets
in abbreviated notation:
\be
\begin{array}{ccccc}
\langle W_{-1} V_\alpha(0) \rangle = & &
- \langle W_{-1} V_{\alpha_3}(1) \rangle&
- \langle W_{-2} V_{\alpha_3}(1) \rangle,&  \\
w =& -w_3& - 2\langle W_{-1} V_{\alpha_3}(1) \rangle&
- \langle W_{-2} V_{\alpha_3}(1) \rangle&- w_4,\\
0 =& -3w_3 &- 3\langle W_{-1} V_{\alpha_3}(1) \rangle&
- \langle W_{-2} V_{\alpha_3}(1) \rangle&
-  \langle W_{-1} V_{\alpha_4}(\infty) \rangle
\end{array}
\label{W-1sys}
\ee
This time, however, the four unknowns can not
be found from these three equations: we can only
exclude terms with $W_{-2}$ obtain pair relations
between the terms with $W_{-1}$:
\be
\boxed{
\langle \underline{W_{-1}V_\alpha(0})\ V_{\alpha_3}(1)\
V_{\alpha_4}(\infty)\rangle\ =
\Big(w+w_3+w_4\Big)
\langle V_\alpha(0)\ V_{\alpha_3}(1)\ V_{\alpha_4}(\infty)\rangle\ +\
\langle V_\alpha(0)\ \underline{W_{-1}V_{\alpha_3}(1)}
\ V_{\alpha_4}(\infty)\rangle,
}
\nn\\
\underline{\langle W_{-1}V_\alpha(0)}
\ V_{\alpha_3}(1)\ V_{\alpha_4}(\infty)\rangle\ =
\Big(2w-w_3+2w_4\Big)
\langle V_\alpha(0)\ V_{\alpha_3}(1)\ V_{\alpha_4}(\infty)\rangle\ -\
\langle V_\alpha(0)\ V_{\alpha_3}(1)\
\underline{W_{-1}V_{\alpha_4}(\infty)}\rangle,
\nn\\
\langle V_\alpha(0)\
\underline{W_{-1}V_{\alpha_3}(1)}\ V_{\alpha_4}(\infty)\rangle\ =
\Big(w-2w_3+w_4\Big)
\langle V_\alpha(0)\ V_{\alpha_3}(1)\ V_{\alpha_4}(\infty)\rangle\ -\
\langle V_\alpha(0)\ V_{\alpha_3}(1)\
\underline{W_{-1}V_{\alpha_4}(\infty)}\rangle
\label{W-1rels}
\ee
Only if an additional constraint like (\ref{Wzc}) is imposed on
any {\it one} of the three states, all correlators can be
unambiguously defined in a universal way: through $\Delta$'s
and $w$'s (or, what is equivalent, through the parameters $\vec\alpha$).
It is the first (boxed) of these formulas that was relevant for
our consideration in s.\ref{agt1}.

The triple vertices of another type $\bar\Gamma$ are
matrix elements rather than correlators (averages),
they also obey a Ward identity, similar to (\ref{Lncont}),
but not exactly the same. We refer to \cite{MMMM} for
details of the derivation and present here only the
answers, which we need the most:
\be
\boxed{\bar\Gamma(L_{-1})_{\alpha_1\alpha_2;\alpha} =\
\langle \underline{L_{-1}V_\alpha}| V_{\alpha_1}(1)\
V_{\alpha_2}(0)\rangle\ =
\Big(\Delta_{\alpha}  + \Delta_1  - \Delta_2\Big)
\langle V_{\alpha}\ |\ V_1(1)V_2(0) \rangle,
} \nn \\
\langle V_{\alpha}\ |\
\underline{(L_{-1}V_1)(1)}\ V_2(0)\rangle \ =
\Big(\Delta_{\alpha} - \Delta_1-\Delta_2\Big)
\langle V_{\alpha}\ |\ V_1(1) V_2(0)\rangle,
\label{barGL}
\ee
\vspace{-0.0cm}
$$
\boxed{ \bar\Gamma(W_{-1})_{\alpha_1\alpha_2;\alpha} =\
\langle \underline{W_{-1}V_\alpha} | V_{\alpha_1}(1)\
V_{\alpha_2}(0)\rangle\
= \left(w_{\alpha}+ {2}w_1-w_2\right)
\langle V_{\alpha} | V_1(1)\ V_2(0) \rangle\  +
\langle V_{\alpha}  |
\underline{(W_{-1}V_1)(1)}\ V_2(0)\rangle
}
$$
The second formula in (\ref{barGL}) is used to handle the
situation when $V_{\alpha_1}$ is  {\it special}.

\subsection{Shapovalov matrix at level two}

Similar relations at level two are more sophisticated,
but can be straightforwardly derived in the same way.
They are listed in a separate paper \cite{MMMM},
where they are also checked with the help of the
free field model. This check is of importance to make because
the derivation is rather tedious and mistakes are not so
easy to exclude. We present and use these partly-validated
answers in s.5 below.

It remains to evaluate the $W_3$ Shapovalov form.
At level one it is very simple: if expressed in
terms of $D$  from eq.(\ref{Ddef}), it turns from (\ref{SF}) into
\be
\begin{array}{|c|c|}
\hline
2\Delta & 3w \\
\hline
3w & 9D\Delta/2 \\
\hline
\end{array}
\label{SF1}
\ee
for all values of $c$.
At level two there are five "$W_3$-Young diagrams" ${\cal Y}$
with $|{\cal Y}|=2$ and one gets a $5\times 5$ matrix:

\be
\begin{array}{|c||c|c|c|c|c|}
\hline
&&&&&\\
Q_{\alpha}({\cal Y},{\cal Y}')
& L_{-2} V_{\alpha} &  L_{-1}^2 V_{\alpha} & L_{-1}W_{-1} V_{\alpha}
& W_{-2} V_{\alpha} & W_{-1}^2 V_{\alpha} \\
&&&&&\\
\hline\hline
&&&&&\\
L_{-2} V_{\alpha}&4\Delta+\frac{c}{2}&6\Delta&9w&6w&\frac{45D\Delta}{2}\\
&&&&&\\
\hline
&&&&&\\
 L_{-1}^2 V_{\alpha}&6\Delta&4\Delta(2\Delta+1)&6w(2\Delta+1)&12w
 &9(3D\Delta+2w^2)\\
&&&&&\\
\hline
&&&&&\\
L_{-1}W_{-1} V_{\alpha}&9w&6w(2\Delta+1)&9(D\Delta^2 + D\Delta + w^2)
&18D\Delta&\frac{27Dw}{2}(2\Delta+3)\\
&&&&&\\
\hline
&&&&&\\
W_{-2} V_{\alpha}&6w&12w&18D\Delta
& 9\Delta(D+1)&\frac{27w}{2}(3D+1)\\
&&&&&\\
\hline
&&&&&\\
W_{-1}^2 V_{\alpha}&\frac{45D\Delta}{2}&9(3D\Delta+2w^2)
&\frac{27Dw}{2}(2\Delta+3)&\frac{27w}{2}(3D+1)
&\frac{81}{4}
D^2\Delta(2\Delta+1)+\\
&&&&&\frac{81\varkappa}{4} \Big(D\Delta(\Delta+1)+4w^2
\Big)\\
&&&&&\\
\hline
\end{array}
\label{SF2}
\ee
where special notation $\varkappa$ is introduced for a peculiar combination
\be
\varkappa = \frac{32}{22+5c} \stackrel{c=2}{\longrightarrow}\ 1,
\ \ \ \ {\rm then} \ \ \ \
D=\varkappa\left(\Delta+\frac{1}{5}\right) - \frac{1}{5}
\ \stackrel{c=2}{\longrightarrow}\  \Delta
\label{Ddef1}
\ee
We describe here only evaluation of the tricky entries
of (\ref{SF2}), which involves commutators of $W$-operators.
The first such example is
\be
<L_{-2}V_\alpha|W_{-1}^2V_\alpha> =
<\alpha|L_2W_{-1}^2|\alpha> \ \stackrel{(\ref{LWcom})}{=}\
<\alpha|5W_1W_{-1}|\alpha>
\ \stackrel{(\ref{WWcom})}{=}\
\frac{45}{2}D\Delta
\ee
At the last stage we could of course directly use (\ref{WnW-n}).
As in \cite{MMMagt} we use obvious abbreviated notation,
which we now abbreviate even further. Similarly,
\be
L_1^2W_{-1}^2 \ \stackrel{(\ref{LWcom})}{=}\
L_1(W_{-1}L_{-1}+3W_0)W_{-1} =
(L_1W_{-1})^2+3(L_1W_{-1})W_0 + 3L_1\, [W_0,W_{-1}]
\label{form1}
\ee
The first two terms are straightforward: they give
$(3w)^2 + 3(3w)w = 18w^2$ when acting on a primary field.
The last term is trickier: according to (\ref{WWcom}),
the commutator turns into
$\frac{9}{2}\Big((\varkappa/2) \Lambda_{-1} - (1/5)L_{-1}\Big)$
and then
into $\frac{9}{2}L_{-1}\Big(\varkappa(\Delta + 1/5)-1/5\Big)
= \frac{9D}{2}L_{-1}$. Finally,
$3L_1L_{-1} \rightarrow 6\Delta$ and one obtains for this
entry of the table: $18w^2 + 27D\Delta$.
Alternatively, in (\ref{form1}) one could push $W_0$ to the
left: $3L_{1}W_0W_{-1} = 3(W_0L_1+2W_1)W_{-1} \rightarrow
9w^2 + 27D\Delta$, which reproduces the same result in a simpler way.

The other entries of the matrix (\ref{SF2}) are calculated
in a similar way:
\be
W_1L_1L_{-1}W_{-1} = W_1(2L_0+L_{-1}L_{1})W_{-1} \rightarrow
2W_1\Big(W_{-1}(L_0+1)\Big) + 3W_1L_{-1}W_0 \rightarrow
9D\Delta(\Delta+1) + 9w^2;
\ee
\be
W_1L_1W_{-2} \rightarrow 4W_1W_{-1} \rightarrow 18D\Delta;
\ee
\be
W_1L_{1}W_{-1}^2 \rightarrow W_1(W_{-1}L_1+3W_0)W_{-1}
\rightarrow \frac{27}{2}wD\Delta + \frac{27}{2}wD\Delta +
\frac{27D}{2}W_1L_{-1} \rightarrow \frac{27wD}{2}(2\Delta+3)
\ee
\be
W_1^2W_{-2} \rightarrow W_1 [W_1,W_{-2}] =
\frac{27}{2}W_1
\left(\frac{\varkappa}{2}\Lambda_{-1} + \frac{2}{15}L_{-1}\right)
\rightarrow \frac{27}{2}W_1L_{-1}
\left(\varkappa\Big(\Delta+\frac{1}{5}\Big)+\frac{2}{15}\right)
\rightarrow \frac{81w}{2}\left(D+\frac{1}{3}\right)
\ee
\be
W_1^2W_{-1}^2 = W_1\left(W_{-1}W_1 + \frac{9}{2}
\Big({\varkappa}\Lambda_0 - \frac{1}{5}L_0\Big)\right)W_{-1}
\rightarrow \left(\frac{9D\Delta}{2}\right)^2
+\frac{9\varkappa}{2}W_1\Lambda_0W_{-1} -
\frac{9}{10}\cdot\frac{9}{2}D\Delta(\Delta+1) \rightarrow
\ee
The middle term in the last expression is a little tedious: since $\Lambda_0$
acts on a first descendant rather than primary,
one needs to include
also a term from the normal ordered part:
$W_1\Lambda_0W_{-1} \rightarrow
W_1\left(L_0^2+\frac{1}{5}L_0+2L_{-1}L_1\right)W_{-1}
\rightarrow \frac{9D\Delta}{2}(\Delta+\frac{6}{5})(\Delta+1)
+18w^2$. Collecting all terms, one finally gets
\be
W_1^2W_{-1}^2 \rightarrow \frac{81}{4}\left(
D^2\Delta(2\Delta+1)+\varkappa \big(D\Delta(\Delta+1)+4w^2
\big)\right)
\ee

\section{
AGT relation to $c\neq 2$ at level one
\label{cn2}}

As was already mentioned, going to arbitrary central charges
$c$ and to higher levels, i.e. to higher powers of $x$ and bigger
sizes of the generalized Young diagrams,
is a straightforward, but tedious exercise in group
theory. The crucial difficult fact is actually well known:
the Kac determinant is {\it always} (i.e. for arbitrary level
and for arbitrary $W_N$ algebra) nicely factorized in terms of
$\vec\alpha$-variables, so that its zeroes are always given
by an integer combination
$\vec\alpha \vec e_i= m\epsilon_1 + n\epsilon_2$.
This fact is remarkable, but well known, its best
heuristic "explanation" comes from the theory of free fields
\cite{fref}, but algebraically it looks somewhat artificial.

\subsection{No external lines, $\vec\alpha_1,\ldots,\vec\alpha_4=0$
\label{lev1noext}}

At level one for arbitrary $c$ it follows from existence
of the deformation of (\ref{va}):
\be
\Delta^3-w^2 =
(\alpha^2+\beta^2)^3 - \Big(\alpha(\alpha^2-3\beta^2)\Big)^2 =
 \Big(\beta(\beta^2-3\alpha^2)\Big)^2 \ \longrightarrow \
\nn \\
\boxed{
(\alpha^2+\beta^2-Q^2)^2\left(\alpha^2+\beta^2-\frac{1}{4}Q^2\right) -
\alpha^2(\alpha^2-3\beta^2)^2 =
(\beta^2-Q^2/4)\Big((\beta-Q)^2 - 3\alpha^2\Big)
\Big((\beta+Q)^2-3\alpha^2\Big)
}
\label{defDeltaw}
\ee
The l.h.s. of this deformed relation is
interpreted in accordance with (\ref{SF1}), as determinant of the
Shapovalov matrix,
\be
D\Delta^2-w^2 =
\left(\frac{1}{1-\frac{15}{4}Q^2}(\alpha^2+\beta^2-Q^2+\frac{1}{5})
- \frac{1}{5}\right)(\alpha^2+\beta^2-Q^2)^2 -
\frac{1}{1-\frac{15}{4}Q^2}\alpha^2(\alpha^2-3\beta^2)^2 = \nn \\
=\frac{1}{1-\frac{15}{4}Q^2}\left\{
(\alpha^2+\beta^2-Q^2)^2\left(\alpha^2+\beta^2-\frac{1}{4}Q^2
\right) -
\alpha^2(\alpha^2-3\beta^2)^2\right\}
\ee
i.e. $c = 2(1-12Q^2)$ and $\Delta=\alpha^2+\beta^2-Q^2$.
Note that the eigenvalue $w$
and, thus, the operator $W^{(3)}$ itself also acquire
the $c$-dependent factor $\sqrt{\frac{32}{22+5c}}$.

In the AGT relation, the r.h.s. of (\ref{defDeltaw})
should match with a deformation of square of the Van-der-Monde
determinant in the Nekrasov formula, given by the denominator
in eq.(\ref{Z1gen}): instead of (\ref{vVDMa}) one now has
\be
(\beta^2-Q^2/4)\Big((\beta-Q)^2 - 3\alpha^2\Big)
\Big((\beta+Q)^2-3\alpha^2\Big) \sim \nn \\ \sim
(a_{12}-\epsilon)a_{12}(a_{12}+\epsilon)
(a_{23}-\epsilon)a_{23}(a_{23}+\epsilon)
(a_{31}-\epsilon)a_{31}(a_{31}+\epsilon)
\label{dencneq2}
\ee
Note that the number of different structures at the r.h.s.
(\ref{dencneq2}), i.e. in the denominator
of the Nekrasov formula, increased from six to nine with switching
on a non-zero $\epsilon$.
Some three zeroes of the denominator should be canceled
by the numerator of the Nekrasov formula, and only the remaining
six should
match the six zeroes at the l.h.s. of (\ref{dencneq2}).
From experience with the $SU(2)$ case in \cite{MMMagt},
one can guess that {\it irrelevant} three factors at the
r.h.s. of (\ref{dencneq2})
are nothing but $a_{12}a_{23}a_{31}=\Delta(a)$,
and the mapping
(\ref{avsalphabeta}) should be deformed so that
\be
\left(\begin{array}{c}
a_{12} = -2\beta \\
a_{23} = \beta+\alpha\sqrt{3}\\
a_{31} = \beta-\alpha\sqrt{3}
\end{array}\right)
\ \longrightarrow \
\left(\begin{array}{c}
a_{12}+\epsilon = -2\beta + Q \\
a_{23}+\epsilon = \beta+Q+\alpha\sqrt{3}\\
a_{31}+\epsilon = \beta+Q-\alpha\sqrt{3}
\end{array}\right)
\ \ \ {\rm and} \ \ \
\left(\begin{array}{c}
a_{12}-\epsilon = -2\beta - Q \\
a_{23}-\epsilon = \beta-Q+\alpha\sqrt{3}\\
a_{31}-\epsilon = \beta-Q-\alpha\sqrt{3}
\end{array}\right)
\ee
i.e. $\epsilon=Q$.
In fact, after appropriate rescalings, see \cite{MMMagt},
\be
Q^2=\frac{\epsilon^2}{-\epsilon_1\epsilon_2}
\ee
and
\be
\Delta_{\vec\alpha} = \frac{\alpha^2+\beta^2-Q^2}{-\epsilon_1\epsilon_2},
\nn \\
w_{\vec\alpha} = \sqrt{\frac{\varkappa}{-\epsilon_1\epsilon_2}}
\,\alpha(\alpha^2-3\beta^2)
\ee
Note that (\ref{avsalphabeta}) actually remains intact:
\be
a_1 = \kappa (\alpha-\beta\sqrt{3}), \nn \\
a_2 = \kappa (\alpha+\beta\sqrt{3}), \nn \\
a_3 = -2\kappa\alpha
\label{avsalphabetacneq2}
\ee
with $\kappa = 1/\sqrt{3}$.

For generic $c$ the AGT relation (\ref{ZB1}) turns into
\be
\frac{1}{2}\cdot
\frac{\Delta\left(D\Delta^2-\frac{8}{9}w^2\right)}
{D\Delta^2-w^2} =\phantom{fgakfgafgafgasfgafgafgafg}\nn \\ =
\frac{1}{\epsilon_1\epsilon_2}\left(
\frac{P_6(a_1)}{a_{12}(a_{12}+\epsilon)a_{13}(a_{13}+\epsilon)} +
\frac{P_6(a_2)}{a_{21}(a_{21}+\epsilon)a_{23}(a_{23}+\epsilon)} +
\frac{P_6(a_3)}{a_{31}(a_{31}+\epsilon)a_{32}(a_{32}+\epsilon)}
\right) + \nu
\label{ZB1c}
\ee
where the conformal block at the l.h.s. is the deformation
of (\ref{B1c2}) and the r.h.s. is the level one
Nekrasov function for
non-vanishing $\epsilon=\epsilon_1+\epsilon_2$.
The matching of denominators is guaranteed by the basic
relation (\ref{defDeltaw}), to which we adjusted all our choices
of dimensions and the AGT relation
$\vec a\ \stackrel{(\ref{avsalphabetacneq2})}{\leftrightarrow}\
\vec\alpha$. Therefore (\ref{ZB1c}) defines
the polynomial $P(a)$ and the $U(1)$ parameter $\nu$.
Like in the $SU(2)$ case in \cite{MMMagt},
some $\mu$-parameters are non-vanishing when $e\neq 0$,
even despite the external lines are neglected, therefore, one should not expect
the case under consideration makes any sense. However, it turns out in this case still
there is a solution of the AGT relation with $\nu\neq 0$:
\be
P(a) = a^2\left(a^4+3e a^3 + \frac{13e^2}{4}a^2
+\frac{3e^3}{2}a + \frac{e^4}{4}\right)
\label{Pfree}
\ee
\be\label{84}
\nu = -\frac{3e^2}{4(-\epsilon_1\epsilon_2)}
\ee
This polynomial, however, does not correspond to any $\mu$'s.

\subsection{Introducing external states
\label{muvecs}}

Now we need to switch on non-vanishing $\vec\alpha_1,\ldots,
\vec\alpha_4$, and impose the speciality conditions on
$\vec\alpha_1$ and $\vec\alpha_3$, this will modify
$B^{(1)}$ at the l.h.s. of (\ref{ZB1c}). Then we use
this relation to find the modified polynomial $P(a)$,
its roots. Surprisingly or not, they will indeed be just
linear functions of $\alpha$'s and $\beta$'s, as
required by the AGT conjecture, and they will provide us
with the deformation of (\ref{mu1c2gen})
and (\ref{mu1c2genvect}).

In what follows we omit the normalization factor
$-\epsilon_1\epsilon_2$ from most formulas. It can be
always restored on dimensional grounds.

The {\it speciality} condition changes:
it is defined by the zeroes of the r.h.s. of (\ref{defDeltaw}),
and therefore we have six options:
\be\label{ssc11}
{\rm either}\ \ \ \
\beta_1 = \pm Q/2 \ \ \ \ {\rm or}\ \ \
\beta_1 = \pm\alpha_1\sqrt{3} \pm Q
\ee
for such vectors $\frac{w_1}{\Delta_1} = \sqrt{D_1}$,
thus when this ratio appears in our formulas,
it does not produce new poles.
These six options are of course related by Weyl group
transformations. However, since we have two special
states in a $4$-point conformal block, there is
a freedom to choose their relative orientation.

Note that the AGT relation itself is correct for all
possible choices in (\ref{ssc11}). However, if fact, not all of them correspond to
the special states (\ref{Wzc}). Indeed, (\ref{Wzc}) is more restrictive than just the
condition of zero Kac determinant (\ref{defDeltaw}). Of 6 states (\ref{ssc11}) only
3 actually correspond to (\ref{Wzc}), \cite{MMMM}
\be\label{ssc1}
{\rm either}\ \ \ \
\beta_1 = - Q/2 \ \ \ \ {\rm or}\ \ \
\beta_1 = \pm\alpha_1\sqrt{3} - Q
\ee
Now all 9 possible combinations of choosing special states $\vec\alpha_{1,3}$
should be considered. For instance, instead of (\ref{mu1c2gen}),
\be
\begin{array}{crcc}
\mu_1=&\frac{2}{\sqrt{3}}(\alpha_2-\alpha_1)
=& \frac{2}{\sqrt{3}}\alpha_2
&-\frac{2\alpha_1}{\sqrt{3}}, \\
\mu_2=&-\frac{\alpha_2+2\alpha_1+\beta_2\sqrt{3}}{\sqrt{3}}
=& -\frac{\alpha_2+\beta_2\sqrt{3}}{\sqrt{3}}
&-\frac{2\alpha_1}{\sqrt{3}}, \\
\mu_3=&-\frac{\alpha_2+2\alpha_1-\beta_2\sqrt{3}}{\sqrt{3}}
=&-\frac{\alpha_2-\beta_2\sqrt{3}}{\sqrt{3}}
&-\frac{2\alpha_1}{\sqrt{3}}, \\
\mu_4=&\frac{2}{\sqrt{3}}(\alpha_4-\alpha_3)
=&\frac{2}{\sqrt{3}}\alpha_4
&-\frac{2\alpha_3}{\sqrt{3}}, \\
\mu_5=&-\frac{\alpha_4+2\alpha_3+\beta_4\sqrt{3}}{\sqrt{3}}
=&-\frac{\alpha_4+\beta_4\sqrt{3}}{\sqrt{3}}
&-\frac{2\alpha_3}{\sqrt{3}}, \\
\mu_6=&-\frac{\alpha_4+2\alpha_3-\beta_4\sqrt{3}}{\sqrt{3}}
=&-\frac{\alpha_4-\beta_4\sqrt{3}}{\sqrt{3}}
&-\frac{2\alpha_3}{\sqrt{3}}
\end{array}
\label{mu1c2gen0}
\ee
and
\be
\nu = -4\alpha_1\alpha_3
\ee
one has, e.g., for

\bigskip

\underline{$\beta_1 =-\epsilon/2,\ \beta_3=-\epsilon/2:$}

\be\begin{array}{crcc}
\mu_1=&\frac{2\alpha_2-\alpha_1}{\sqrt{3}}+\frac{1}{2}\epsilon
=& \frac{2}{\sqrt{3}}\alpha_2
&+ \frac{1}{\sqrt{3}}\left(-\alpha_1+\frac{\sqrt{3}}{2}\epsilon\right), \\
\mu_2=&-\frac{\alpha_2+\alpha_1+\beta_2\sqrt{3}}{\sqrt{3}}+\frac{1}{2}\epsilon
=& -\frac{\alpha_2+\beta_2\sqrt{3}}{\sqrt{3}}
&+ \frac{1}{\sqrt{3}}\left(-\alpha_1+\frac{\sqrt{3}}{2}\epsilon\right), \\
\mu_3=&-\frac{\alpha_2+\alpha_1-\beta_2\sqrt{3}}{\sqrt{3}} +\frac{1}{2}\epsilon
=&-\frac{\alpha_2-\beta_2\sqrt{3}}{\sqrt{3}}
&+ \frac{1}{\sqrt{3}}\left(-\alpha_1+\frac{\sqrt{3}}{2}\epsilon\right), \\
\mu_4=&\frac{2\alpha_4+\alpha_3}{\sqrt{3}}+\frac{1}{2}\epsilon
=&\frac{2}{\sqrt{3}}\alpha_4
&+ \frac{1}{\sqrt{3}}\left(\alpha_3+\frac{\sqrt{3}}{2}\epsilon\right), \\
\mu_5=&-\frac{\alpha_4-\alpha_3+\beta_4\sqrt{3}}{\sqrt{3}}+\frac{1}{2}\epsilon
=&-\frac{\alpha_4+\beta_4\sqrt{3}}{\sqrt{3}}
&+ \frac{1}{\sqrt{3}}\left(\alpha_3+\frac{\sqrt{3}}{2}\epsilon\right), \\
\mu_6=&-\frac{\alpha_4-\alpha_3-\beta_4\sqrt{3}}{\sqrt{3}}+\frac{1}{2}\epsilon
=&-\frac{\alpha_4-\beta_4\sqrt{3}}{\sqrt{3}}
&+\frac{1}{\sqrt{3}}\left(\alpha_3+\frac{\sqrt{3}}{2}\epsilon\right)
\end{array}
\label{mu1c2gen1-1}
\ee
with
\be
\nu=  - \left(-\alpha_1+\frac{\sqrt{3}}{2}e\right)
\left(\alpha_3+\frac{\sqrt{3}}{2}e\right)
\ee
Putting $\alpha_1=0$ and $\alpha_3=0$ here, one
reproduces the solution (\ref{84}).

Other choices of the special states preserve this structure of answer for $\mu$'s.
That is, they are equal to the sum of the same first terms as in (\ref{mu1c2gen1-1})
plus some quantities $S(\alpha_{1,3})$ which depends on the choice of the special
states:
\be\label{genmu}
\begin{array}{ccc}
\mu_1=\frac{2}{\sqrt{3}}\alpha_2+S(\vec\alpha_1),
&& \mu_4=-\frac{2}{\sqrt{3}}\alpha_4+S(\vec\alpha_3), \\
\mu_2=-\frac{1}{\sqrt{3}}(\alpha_2+\beta_2\sqrt{3})+S(\vec\alpha_1),
&& \mu_5=\frac{1}{\sqrt{3}}(\alpha_4+\beta_4\sqrt{3})+S(\vec\alpha_3),\\
\mu_3=-\frac{1}{\sqrt{3}}(\alpha_2-\beta_2\sqrt{3})+S(\vec\alpha_1),
&& \mu_6=\frac{1}{\sqrt{3}}(\alpha_4-\beta_4\sqrt{3})+S(\vec\alpha_3)
\end{array}
\ee
with
\be\label{genS}
S(\alpha_1,-\epsilon/2) = -\frac{1}{\sqrt{3}}\alpha_1 + \frac{1}{2}\epsilon, \ \ \
S\Big(\alpha_1,-\alpha_1\sqrt{3} -\epsilon\Big)
= \frac{2}{\sqrt{3}}\alpha_1+\epsilon, \ \ \
S\Big(\alpha_1,\alpha_1\sqrt{3} - \epsilon\Big)
= \frac{2}{\sqrt{3}}\alpha_1\nn
\\
S(\alpha_3,-\epsilon/2) = \frac{1}{\sqrt{3}}\alpha_3 +
\frac{1}{2}\epsilon, \ \ \ S\Big(\alpha_3,-\alpha_3\sqrt{3} -
\epsilon\Big) = -\frac{2}{\sqrt{3}}\alpha_3, \ \ \
S\Big(\alpha_3,\alpha_3\sqrt{3} - \epsilon\Big) =
-\frac{2}{\sqrt{3}}\alpha_3+\epsilon \ee and \be\label{gennu}
\nu=-3S(\alpha_1)S(\alpha_3) \ee The manifest formulas for
$S(\alpha_{1,3})$ for all possible choices of the special states are
collected in the Conclusion.

\section{Calculations at level two
\label{secl}}

\subsection{Level-two Nekrasov functions}

Nekrasov functions for $SU(N)$ with $N_f=2N$ fundamentals
-- the ones appearing in the AGT representation
of the $4$-point conformal block with $2$ special external states --
are labeled by $N$-plets of ordinary Young diagrams
$\vec Y = \{Y_1,\ldots,Y_N\}$ and are given by \cite{Nek}:
\be
Z_{\vec Y} =
\frac{\prod_{i=1}^N \prod_{\Box\in Y_i} \prod_{f=1}^{N_f}
\Big(\phi(a_i,Y_i,\Box) + \mu_f\Big)}
{\prod_{i,j=1}^N \prod_{\Box\in Y_i}
E(a_i-a_j,Y_i,Y_j,\Box)\Big(\epsilon - E(a_i-a_j,Y_i,Y_j,\Box)\Big)}
\ee
where for a box with coordinates $(m,n)$ in the Young diagram
\be
\phi(a,Y,\Box_{m,n}) = a + \epsilon_1(m-1) + \epsilon_2(n-1),\nn\\
E(a,Y_1,Y_2,\Box_{m,n}) = a + \epsilon_1\Big(k^T_n(Y_1)-m+1\Big)
- \epsilon_2\Big(k_m(Y_2)-n\Big)
\ee
$k^T_n(Y)$ and $k_m(Y)$ are the length of the $n$-th row and
the height of the $m$-th column in the diagram $Y$.

If $Y_2=Y_1$, then at $\epsilon=0$ the combination
$h(\Box_{m,n}) = E(0,Y,Y)=\big(k^T_n(Y)-m+1\big) + (\big(k_m(Y)-n\big)$
is the hook length for the box $(m,n)$, which enters, for example,
the celebrated hook formula for
\be
d_Y = \prod_{\Box\in Y} h^{-1}(\Box)
\ee
for the character $\chi_Y(t_k=\delta_{k,1})=d_Y$, which defines
up to various simple factors dimensions of representations
of symmetric and linear groups.
Thus, for $\epsilon=0$ the Nekrasov
formulas are more or less natural objects in the theory
of character expansions \cite{charex,chardeco} and Hurwitz-Kontsevich
partition functions
\cite{HuKo}, while for $\epsilon\neq 0$ they belong to
representation theory of more sophisticated (quantum affine?)
algebras. The AGT relation associates the deformation to $\epsilon\neq 0$
with the deformation to $c\neq r=N-1$ in the theory of
$r$ free fields (conformal Toda?).
The square $d^2_Y$ is also known as the Plancherel measure for
integer partitions, which appears in the theory of KP and Kontsevich-Hurwitz
$\tau$-functions, beginning from the famous sum rule
\be
\sum_Y \chi_Y(t)\chi_Y(t') = \exp\left(\sum_k kt_kt'_k\right)
\ee
and its particular case
\be
\sum_Y d^2_Y x^{|Y|} = e^x
\ee
The Nekrasov formulas with $\epsilon\neq 0$
provide a deformation of this measure.

At level one, only one kind of $N$-plets $\vec Y$ contributes,
containing just one diagram of the unit size:
\be
Z_{[1]} = -\frac{1}{\epsilon_1\epsilon_2}\sum_{i=1}^N
\frac{P(a_i)}{\prod_{k\neq i}^N a_{ik}(a_{ik}+\epsilon)}
\equiv -\frac{1}{\epsilon_1\epsilon_2}\sum_{i=1}^N R_i(a_i)
\ee
At level two, only three kinds of $N$-plets $\vec Y$ contribute:
containing just one diagram of size two, either $[2]$ or $[11]$,
or containing a pair of the single-box $[1]$ diagrams.
The corresponding Nekrasov functions are
\be
Z_{[2]} = \frac{1}{2!\epsilon_1\epsilon_2^2(\epsilon_1-\epsilon_2)}
\sum_{i=1}^N R_i(a_i)R_i(a_i+\epsilon_2),\nn\\
Z_{[11]} = -\frac{1}{2!\epsilon_1^2\epsilon_2(\epsilon_1-\epsilon_2)}
\sum_{i=1}^N R_i(a_i)R_i(a_i+\epsilon_1),\nn\\
Z_{[1][1]} = \frac{1}{\epsilon_1^2\epsilon_2^2}\sum_{i< j}^N
R_i(a_i)R_j(a_j){a_{ij}^2 (a_{ij}^2-\epsilon^2)\over (a_{ij}^2-\epsilon_1^2)
(a_{ij}^2-\epsilon_2^2)}
\label{Nek2}
\ee
See also eq.(62) of \cite{MMMagt} for a particular case of $N=2$.

\subsection{AGT relations at level two}

The AGT relation at level two states that
\be
B^{(2)} = Z_{[2]}+Z_{[11]} + Z_{[1][1]} +\nu Z_{[1]} + \frac{\nu(\nu+1)}{2}
\label{ZB2}
\ee
where $Z_{[1]} = Z_1$.
The denominator at the l.h.s. comes from inverse of the
Shapovalov form, i.e. equals to the Kac determinant of matrix (\ref{SF2}),
while at the r.h.s. it contains the product of $a_{ij}$, shifted by
linear combinations of $\epsilon_1$ and $\epsilon_2$.
In other words, the AGT relation implies, to begin with, the
matching of the denominators, which requires that
\be
{\rm Kac\ determinant}\ =
\det \Big({\rm Shapovalov\ matrix}\Big) \sim
\prod_{i\neq j} \prod_{p,q > 0}
(a_{ij}+p\epsilon_1+q\epsilon_2)
\label{KDfact}
\ee
The fact that the Kac determinants factorize in such an elegant
way, if  expressed through the $\vec\alpha$-variables instead
of the eigenvalues $\Delta$ and $w$ of the $W$-operators,
is one of the central results of representation theory of
the Virasoro and $W$-algebras.
In s.\ref{kdfact} we illustrate this remarkable general result in
the particular example of level two for the $W_3$ algebra.

Note, however, that the sets $\{(p,q)\}$ are actually
not obviously the same at the l.h.s. and the r.h.s. of (\ref{KDfact}).
In the Kac determinant they are constrained by the condition
$0<pq\leq {\rm level}$, while in Nekrasov's formulas at the
same level one can expect that either $p$, or $q$ can be equal to zero.
These extra zeroes of the Nekrasov determinant are actually
canceled between different terms in the sum over Young diagrams,
however, particular Nekrasov functions have {\it more} singularities
than there are zeroes of the Kac determinant.

An additional claim as compared to the Kac determinant factorization property
in the AGT conjecture is that the numerators
also match for an appropriate {\it linear} relation between
$\mu_f$ and $\vec\alpha_{1,2,3,4}$ parameters
(with $\vec\alpha{1,3}$ subjected to the
{\it speciality} constraints).
We do not present details of this check here:
at the present level of understanding
it is more a computer exercise than a conceptual calculation.
Some necessary comments are given in s.\ref{coblo2}.

\subsection{Factorization of Kac determinant
\label{kdfact}}

For $c=2$ the determinants of the level-one and level-two
Shapovalov matrices (\ref{SF}) and (\ref{SF2}) are
equal to
\be
3^2\beta^2(\beta^2-3\alpha^2)^2 \sim \Big(a_{12}a_{23}a_{31}\Big)^2
\label{den1c2}
\ee
and
\be
3^8\beta^4(2\beta-1)^2(2\beta+1)^2
(\beta^2-3\alpha^2)^4
\Big((\beta-1)^2-3\alpha^2\Big)^2
\Big((\beta+1)^2-3\alpha^2\Big)^2 \sim \nn \\
\sim \Big((a_{12}-1)a_{12}^2(a_{12}+1)(a_{23}-1)a_{23}^2(a_{23}+1)
(a_{31}+1)a_{31}^2(a_{31}-1)\Big)^2
\label{den2c2}
\ee
respectively. This is in perfect agreement with (\ref{KDfact}),
provided $\epsilon_1=-\epsilon_2=1$ and
identification (\ref{avsalphabeta}) is made:
\be
a_{12}=-2\beta, \ \ \ a_{23}=\beta+\alpha\sqrt{3}, \ \ \
a_{31} = \beta-\alpha\sqrt{3},\ \ \ \ {\rm i.e.}\nn\\
a_1 = \frac{1}{\sqrt{3}}(\alpha-\beta\sqrt{3}), \ \ \
a_2 = \frac{1}{\sqrt{3}}(\alpha+\beta\sqrt{3}), \ \ \
a_3 = -\frac{2\alpha}{\sqrt{3}}
\label{avsal}
\ee
(the relative normalization of $a$ and $\alpha$ in
(\ref{avsal}) is defined by the full AGT relation
and is not obvious at the level of Kac determinants).

For generic $c$ the degeneration that gives rise to full squares
in the Nekrasov denominators is resolved, and the same happens
to the Kac determinant.  At level one, one obtains instead of
(\ref{den1c2}):
\be
K_1=\frac{3^2}{4-15Q^2}(4\beta^2-Q^2)
\Big((\beta-Q)^2-3\alpha^2\Big)
\Big((\beta+Q)^2-3\alpha^2\Big) \sim \nn\\
\sim (2\beta-Q)(2\beta+Q)
(\beta+\alpha\sqrt{3}-Q)(\beta+\alpha\sqrt{3}+Q)
(\beta-\alpha\sqrt{3}-Q)(\beta-\alpha\sqrt{3}+Q)
\sim \nn \\ \sim
\prod_{i\neq j} (a_{ij}+\epsilon) =
(a_{12}^2-\epsilon^2)(a_{23}^2-\epsilon^2)(a_{31}^2-\epsilon^2)
\label{den1}
\ee
Therefore, identification (\ref{avsal}) remains intact,
one should only add that $Q=\epsilon$.

At level two, the l.h.s. of (\ref{den2c2}) is deformed into
\be
K_2=\frac{2^4\cdot 3^8}{(4-15Q^2)^4}
\left\{(4\beta^2-Q^2)
\Big((\beta-Q)^2-3\alpha^2\Big)
\Big((\beta+Q)^2-3\alpha^2\Big)\right\}^2\cdot\nn\\
(2Q^2-1-6\beta Q+4\beta^2)(2Q^2-1+6\beta Q+4\beta^2)\cdot\nn\\
(1+12\beta Q^3+6\beta^3 Q-6\beta Q-18\alpha^2\beta Q
-15\alpha^2 Q^2-6\alpha^2-2\beta^2-4Q^2-6\alpha^2\beta^2
+13\beta^2Q^2+\beta^4+4Q^4+9\alpha^4)\cdot\nn\\
(1-12\beta Q^3-6\beta^3 Q+6\beta Q+18\alpha^2\beta Q
-15\alpha^2 Q^2-6\alpha^2-2\beta^2-4Q^2-6\alpha^2\beta^2
+13\beta^2 Q^2+\beta^4+4Q^4+9\alpha^4)
\ee
so that the level one Kac determinant factors out from that of level two
for arbitrary
$\Delta$ and $w$, and the rest of the formula
(including the second power of (\ref{den1}))
factorizes nicely in the $\alpha$-parametrization.
The dictionary for decoding this quantity is simple:
for $\epsilon_1+\epsilon_2=\epsilon=Q$ and $\epsilon_1\epsilon_2=-1$
one has
\be
(2\beta+\epsilon_1)(2\beta+\epsilon_2) =
4\beta^2+2Q\beta - 1, \nn\\
(2\beta+2\epsilon_1+\epsilon_2)(2\beta+\epsilon_1+2\epsilon_2)
= 4\beta^2+6Q\beta + 2Q^2-1
\ee
and
\be
(\beta+\alpha\sqrt{3}+\epsilon_1)(\beta+\alpha\sqrt{3}+\epsilon_2)
(\beta-\alpha\sqrt{3}+\epsilon_1)(\beta-\alpha\sqrt{3}+\epsilon_2)
= \nn\\
1-6\alpha^2\beta Q+2Q\beta^3+2Q\beta-6\alpha^2\beta^2-3\alpha^2Q^2
+6\alpha^2+2\beta^2+\beta^2Q^2+9\alpha^4\beta^4, \nn \\ \nn \\
(\beta+\alpha\sqrt{3}+2\epsilon_1+\epsilon_2)
(\beta+\alpha\sqrt{3}+\epsilon_1+2\epsilon_2)
(\beta-\alpha\sqrt{3}+2\epsilon_1+\epsilon_2)
(\beta-\alpha\sqrt{3}+\epsilon_1+2\epsilon_2)
= \nn \\
= (1+12\beta Q^3+6\beta^3 Q-6\beta Q-18\alpha^2\beta Q
-15\alpha^2 Q^2-6\alpha^2-2\beta^2-4Q^2-6\alpha^2\beta^2
+13\beta^2Q^2+\beta^4+4Q^4+9\alpha^4)
\ee
Thus, one can see that
\be
K_2 \sim
(a_{12}^2-\epsilon^2)^2(a_{23}^2-\epsilon^2)^2(a_{31}^2-\epsilon^2)^2
\Big(a_{12}^2-(\epsilon_1+2\epsilon_2)^2\Big)
\Big(a_{23}^2-(\epsilon_1+2\epsilon_2)^2\Big)
\Big(a_{31}^2-(\epsilon_1+2\epsilon_2)^2\Big)\cdot\nn\\ \cdot
\Big(a_{12}^2-(2\epsilon_1+\epsilon_2)^2\Big)
\Big(a_{23}^2-(2\epsilon_1+\epsilon_2)^2\Big)
\Big(a_{31}^2-(2\epsilon_1+\epsilon_2)^2\Big)
\ee
The factors
$(a_{ij}^2-\epsilon_1^2)$ and $(a_{ij}^2-\epsilon_2^2)$,
which one could expect to arise looking at the denominators
in eq.(\ref{Nek2}), do not actually appear in the Kac determinant.
Instead, $K_1$ comes squared.
This means that these factors should cancel out after
summing different terms in the Nekrasov formula
with the appropriately chosen polynomial $P(a)$ in the numerator.

\subsection{Conformal block at level two
\label{coblo2}}

The conformal block $B^{(2)}$ is obtained
from the general expression (\ref{cbexp})
by substitution of the inverted $5\times 5$ Shapovalov matrix (\ref{SF2})
and five pairs triple vertices from \cite{MMMM}.
To define these vertices unambiguously in un-specified
conformal model,
one should impose the {\it speciality} conditions
and make use of (\ref{W-1rels}), what is also done in \cite{MMMM}.

Note that there are five generalized Young diagrams ${\cal Y}$
of size $|{\cal Y}|=2$, two "pure Virasoro", two
"pure $W$" and one "mixed". Starting from level two,
there is no way to separate $W$ and Virasoro diagrams.
Collecting all the seven pairs of vertices for the two first levels
in one place, one gets
\be
\begin{array}{|ll|}
\hline &\\
\bar\Gamma(L_{-1}) =& \Delta+\Delta_1-\Delta_2,  \\ &\\
\bar\Gamma(W_{-1}) =& w+2w_1-w_2 +
\frac{3w_1}{2\Delta_1}(\Delta-\Delta_1-\Delta_2)  \\
&\\ \hline &\\
\bar\Gamma(L_{-2}) =& \Delta +2\Delta_1 -\Delta_2,  \\ &\\
\bar\Gamma(L_{-1}^2) =& \Big(\Delta +\Delta_1 -\Delta_2\Big)
\Big(\Delta +\Delta_1 -\Delta_2+1\Big),\\ &\\
\bar\Gamma(L_{-1}W_{-1}) =& \Big(\Delta+\Delta_1-\Delta_2+1\Big)
\left(w+2w_1-w_2 + \frac{3w_1}{\Delta_1}(\Delta-\Delta_1-\Delta_2)\right),
\\ &\\
\bar\Gamma(W_{-2}) =& w+5w_2-w_2
+ \frac{3w_1}{\Delta_1}(\Delta-\Delta_1-\Delta_2),
\\ &\\
\bar\Gamma(W_{-1}^2) =&
\left(w+2w_1-w_2
+\frac{3w_1}{2\Delta_1}\big(\Delta-\Delta_1-\Delta_2\big)\right)
\left(w+2w_1-w_2
+\frac{3w_1}{2\Delta_1}\big(\Delta-\Delta_1-\Delta_2+1\big)\right)+\\
&\\
& \ \ \ \ \ \ \ \ \ \ \ \ \ \ \ \ \ \ \ \ \ \ \ \ \ \ \ \ \
 \ \ \ \ \ \ \ \ \ \ \ \ \ \ \ \ \ \ \ \ \ \ \ \ \ \ \ \ \
+ \frac{9D}{2}\big(\Delta+\Delta_1-\Delta_2\big)
\\
&\\
\hline
\hline &\\
\Gamma(L_{-1}) =& \Delta+\Delta_3-\Delta_4,  \\ &\\
\Gamma(W_{-1}) =& w+w_3+w_4 -
\frac{3w_3}{2\Delta_3}(\Delta+\Delta_3-\Delta_4)  \\
&\\ \hline &\\
\Gamma(L_{-2}) =& \Delta +2\Delta_3 -\Delta_4,  \\ &\\
\Gamma(L_{-1}^2) =& \Big(\Delta +\Delta_3 -\Delta_4\Big)
\Big(\Delta +\Delta_3 -\Delta_4+1\Big),\\ &\\
\Gamma(L_{-1}W_{-1}) =& \Big(\Delta+\Delta_3-\Delta_4+1\Big)
\left(w+w_3+w_4 - \frac{3w_3}{\Delta_3}(\Delta+\Delta_3-\Delta_4)\right),
\\ &\\
\Gamma(W_{-2}) =& w+w_3+w_4
- \frac{3w_3}{\Delta_3}(\Delta+\Delta_3-\Delta_4),
\\ &\\
\Gamma(W_{-1}^2) =&
\left(w+w_3+w_4
-\frac{3w_3}{2\Delta_3}\big(\Delta+\Delta_3-\Delta_4\big)\right)
\left(w+w_3+w_4
-\frac{3w_3}{2\Delta_3}\big(\Delta+\Delta_3-\Delta_4+1\big)\right) +\\
&\\
& \ \ \ \ \ \ \ \ \ \ \ \ \ \ \ \ \ \ \ \ \ \ \ \ \ \ \ \ \
 \ \ \ \ \ \ \ \ \ \ \ \ \ \ \ \ \ \ \ \ \ \ \ \ \ \ \ \ \
+\frac{9D}{2}\big(\Delta+\Delta_3-\Delta_4\big)\\
&\\
\hline
\end{array}
\label{vertices}
\ee
Here we assumed that $\vec\alpha_1$ and
$\vec\alpha_3$ are the special state, i.e. that
\be
W_{-1}V_{\alpha_1}(1)=\frac{3w_1}{\Delta_1}L_{-1}V_{\alpha_1}(1)
\ \stackrel{(\ref{L-1rels})}{=}
-\frac{3w_1}{2\Delta_1}\Big(\Delta+\Delta_1-\Delta_2\Big)V_{\alpha_1}(1)
\ee
and similarly for $\alpha_3$:
this is why these specific combinations appear in the
formulas. We also omit the common structure constant factors
$C^\alpha_{\alpha_1\alpha_2}
=\langle V_{\alpha}(0)| V_{\alpha_1}(1)\ V_{\alpha_2}(\infty)\rangle$
and
$C_{\alpha\alpha_3\alpha_4}
=\langle V_{\alpha}(0)\ V_{\alpha_1}(1)\ V_{\alpha_2}(\infty)\rangle$
at the r.h.s.
Note the delicate and seemingly irregular sign differences
between formulas for $\bar\Gamma$ and $\Gamma$ vertices,
all these details being essential, since the AGT relations are very sensitive
to details.

For making up a conformal block, one also needs to invert the
Shapovalov matrices (\ref{SF1}) and (\ref{SF2}). For (\ref{SF1})
it is simple:
\be
\frac{1}{K_1}\begin{array}{|c|c|}
\hline
9D\Delta/2& -3w \\
\hline
-3w & 2\Delta  \\
\hline
\end{array}
\label{SF11}
\ee
but this paper is too short to explicitly write down
the inverse of (\ref{SF2}), neither in $\Delta,w$, nor
in $\alpha,\beta$ variables.
Still, multiplying this inverse by the two
$5$-vector $\Gamma$ and $\bar\Gamma$
from (\ref{vertices}) accordingly to
(\ref{cbexp}), one obtains the conformal block $B^{(2)}$
which, indeed, coincides with the combination of $Z^{(2)}$
and $Z^{(1)}$ at the r.h.s. of (\ref{ZB2}), provided
the six $\mu$'s and $\nu$ are given by relations
(\ref{genmu})-(\ref{gennu})
for various choices of {\it special} states
$\vec\alpha_1$ and $\vec\alpha_3$,
which we already found explicitly in the analysis of the first level.
The calculation is tedious, it necessarily includes
the check of expressions (\ref{vertices}) for the $3$-point
functions.
As usual, it is worth starting from the simplest case of $c=2$,
$\vec\alpha_1=\ldots=\vec\alpha_4=0$ (for $c=2$ this does
not contradict {\it speciality} conditions).
In this case, (\ref{vertices}) is simplified to
\be
\begin{array}{|ll|l|}
\hline &&\\
\Gamma(L_{-1}) =& \Delta &=\Delta,  \\ &&\\
\Gamma(W_{-1}) =& w & = w \\
&&\\ \hline &&\\
\Gamma(L_{-2}) =& \Delta & = \Delta,  \\ &&\\
\Gamma(L_{-1}^2) =& \Delta(\Delta +1) &= \Delta(\Delta +1),\\ &&\\
\Gamma(L_{-1}W_{-1}) =& (\Delta+1)w &= (\Delta+1)w,
\\ &&\\
\Gamma(W_{-2}) =& x_1w & =w,
\\ &&\\
\Gamma(W_{-1}^2) =& x_2\Delta^3+x_3w^2 + x_4\Delta^2 + x_5\Delta
&= w^2+\frac{9}{2}\Delta^2\\
&& \\
\hline
\hline &&\\
\Gamma(L_{-1}) =& \Delta &=\Delta,  \\ &&\\
\Gamma(W_{-1}) =& w & = w \\
&&\\ \hline &&\\
\Gamma(L_{-2}) =& \Delta & = \Delta,  \\ &&\\
\Gamma(L_{-1}^2) =& \Delta(\Delta +1) &= \Delta(\Delta +1),\\ &&\\
\Gamma(L_{-1}W_{-1}) =& (\Delta+1)w &= (\Delta+1)w,
\\ &&\\
\Gamma(W_{-2}) =& y_1w & = w,
\\ &&\\
\Gamma(W_{-1}^2) =& y_2\Delta^3+y_3w^2 + y_4\Delta^2 + y_5\Delta
&= w^2+\frac{9}{2}\Delta^2\\
&& \\
\hline
\end{array}
\label{vert}
\ee
and the Nekrasov formula can be expressed through $\Delta$
and $w$ as follows:
$$
Z^{(2)} = \frac{2}{K_2}\left\{81\Delta(\Delta^3-w^2)
\left(\Delta^3-\frac{8}{9}w^2\right)\right\}^2 +
$$ $$
+ \frac{81}{4K_2}(\Delta^2-w^2)\left\{
\Delta \Big(-1863 \Delta^9 + 2160w^2\Delta^6 + 608w^4\Delta^3 -896w^6\Big)
+  6\Big(81 \Delta^9 + 480w^2\Delta^6 -448 w^4\Delta^3 \Big)\right. +
$$ $$
+ \Delta^2 \Big( 3807\Delta^6 -6696 w^2\Delta^3 + 2704w^4\Big)
+ 4\Delta \Big( -1377\Delta^6 + 1422w^2\Delta^3 -152 w^4\Big)
+ 3\Big(1053 \Delta^6 -960 w^2\Delta^3 + 16w^4\Big) +
$$
\vspace{-0.6cm}
\be
\left.
- 81\Delta(10\Delta-1)\Big( \Delta^3 - \frac{8}{9}w^2\Big)
\right\} =
\ee
\be
= \frac{81\Delta^4(\Delta-1)^4\Big(8\Delta^3+9\Delta^2-6\Delta+1\Big) +
G_2w^2 + G_3w^4 + G_4w^6}{4\cdot 81 \cdot (\Delta^3-w^2)
\Big( 4\Delta(\Delta-1)^2 -4w^2\Big)^2},\nn \\
G_2 = -72\Delta(\Delta-1)^2
(25\Delta^5+20\Delta^4-25\Delta^3+23\Delta^2-8\Delta+1), \nn \\
G_4 = 1664\Delta^5+608\Delta^4-2688\Delta^3+2704\Delta^2-608\Delta + 48,\nn\\
G_6 = -128\Delta(4\Delta+7),
\ee
where we grouped terms with the same power of $-\epsilon_1\epsilon_2$
(which is suppressed for sake of space, but can be restored in the formulas)
or, alternatively,
with the same power of $w$.

Eq.(\ref{ZB2}) with $P(a)=a^6$, $\nu=0$
and AGT relation (\ref{avsalphabeta})
between $\vec a$ and $\vec\alpha$ can be {\it used} to find
the ten free parameters $x_{1},\ldots,y_5$.
It is a strongly over-defined system of equations, still
it has a solution:
$x_1=1$, $x_2=0$, $x_3=1$, $x_4 = 9/2$, $x_5= 0$,
$y_1=1$, $y_2=0$, $y_3=1$, $y_4 = 9/2$, $y_5= 0$,
as shown in the right column of the table (\ref{vert}).
Then, validation of the AGT relation at this level comes from
comparison of {\it such} $x$'s and $y$'s with their values in
the CFT table (\ref{vertices}).
After that, one switches on the external states
$\vec\alpha_1,\ldots,\vec\alpha_4$ and finally
deforms to $c\neq 2$.
{\bf This completes the explicit check of the $W_3$ AGT
relation at level two.}

\section{On complete proof of AGT relation in a {\it very special} case
\label{FLpro}}

In \cite{mmNF}, among other things, we provided a proof of the
AGT relation in a very restricted setting.
The idea (also mentioned in \cite{Wyl})
is to make use of exact knowledge of the 4-point conformal block
in the $sl(N)$ Toda theory \cite{FLit} for the {\it very special} kinematics:
when one of external lines is {\it special} and another one is
further restricted to belong to the {\it fully degenerate}
$W$-Verma module.
In $W^{(N)}$ case {\it special} means that $N-2$ conditions
of the type (\ref{Wzc}) are imposed on the $(N-1)$-component
momentum, let it be $\vec\alpha_3$, while {\it fully degenerate}
means that the last remaining component of the {\it special}
$\vec\alpha_1$ is further fixed to a
certain value. Selection rules of the Toda theory
expresses the internal-state momentum $\vec\alpha$ through
$\vec\alpha_1$ and $\vec\alpha_2$, much similarly to the free field model
rule $\vec\alpha = \vec\alpha_1+ \vec\alpha_2$. The only difference is that, in the
Toda case, there are not single, but $N$ possible values of $\vec\alpha$ (i.e.
$N$ different non-zero conformal blocks).
Thus, this Fateev-Litvinov conformal block depends on
$2(N-1) + 1 + 1$ free parameters
($\vec\alpha_2$, $\alpha_3$, $\vec\alpha_4$,$c$)
and on this high-codimension subspace in the total moduli space
of the $4$-point conformal blocks it is represented by
{\it generic} hypergeometric series
\be
\phantom._N F_{N-1}(A_1,\ldots,A_N;B_1,\ldots,B_{N-1},x)
=  1 + x\cdot\frac{A_1\ldots A_N}{B_1\ldots B_{N-1}} +
\frac{x^2}{2}\cdot\frac{A_1(A_1+1)\ldots A_N(A_N+1)}{B_1(B_1+1)\ldots
B_NB_{N+1}} + \nn \\
+ \sum_{n=3}^\infty \frac{x^n}{n!}\cdot
\prod_{k=0}^{n-1}\frac{(A_1+k)\ldots (A_N+k)}{(B_1+k)\ldots (B_{N-1}+k)}
\label{hyse}
\ee
where $2N-1$ parameters $A_i,B_i$ are linear combinations of
$\vec\alpha_2,\alpha_3,\vec\alpha_4$ with coefficients,
depending on the central charge (actually, on
the screening-charge parameters $\epsilon_1$ and $\epsilon_2$).
This remarkable result demonstrates how generic hypergeometric
series are embedded, as a linear subspace if the $\alpha$-parametrization
of this moduli space is used!,
into the space of conformal blocks,
thus, generalizing the old description \cite{MV} of their embedding
into the space of Dotsenko-Fateev integrals, or of hypergeometric
integrals/correlators in the terminology of \cite{SheVa}
(we remind that arbitrary $\phantom._NF_{N-m}$ are obtained
in certain limits from $\phantom._NF_{N-1}$ and lie on
the boundary of the moduli space, see, for example, \cite{MV}).
As explained in \cite{mmNF}, in these terms the AGT conjecture
states that {\bf the Nekrasov functions  provide the generalization of
the $N=2$ hypergeometric series, exactly extending it to
entire space of Virasoro conformal blocks}.
For $N>2$ they are probably sufficient only to extend hypergeometric
series to the moduli space of conformal blocks, restricted to
{\it special} subspace (i.e. when only two out of $m$ external
momenta and all the $m-3$ internal momenta in the
$m$-point conformal block are arbitrary, while the other $m-2$
external momenta belong to $1$-dimensional {\it special} subspaces).
If this is true, there remains a question,
{\bf what provides the further extension beyond Nekrasov functions(?)},
which still remains to be answered.

However, if one reverse the question: what are hypergeometric
series from the point of view of Nekrasov functions,
one immediately arrives to complete proof of the AGT conjecture
for this very restricted (still huge!) class of conformal blocks.
The point is that the answer is very simple:
the hypergeometric series arise when only very specific $N$-plets of
Young diagrams are non-vanishing:
$\vec Y = \Big\{\emptyset,\ldots,[1^n],\ldots,\emptyset\Big\}$
or $\vec Y = \Big\{\emptyset,\ldots,[n],\ldots,\emptyset\Big\}$,
i.e. only one diagram in the $N$-plet is not empty, and it is
either a row $[1^n]$ or column $[n]$.
We call such Nekrasov functions {\it chiral} and {\it anti-chiral} respectively,
explicitly they are
\be
Z(\emptyset,\ldots,[1^n],\ldots,\emptyset) =
\frac{1}{\epsilon_2\epsilon_1^nn!}
\frac{P(a_i)P(a_i+\epsilon_1)\ldots P\big(a_i+(n-1)\epsilon_1\big)}
{(-\epsilon_2)(-\epsilon_2+\epsilon_1)...(-\epsilon_2+(n-1)\epsilon_1)
Q_i(a_i)Q_i\big(a_i+\epsilon_1\big)\ldots
Q_i\big(a_i+(n-1)\epsilon_1\big)}
\ee
and
\be
Z(\emptyset,\ldots,[n],\ldots,\emptyset) = Z(\emptyset,\ldots,[1^n],\ldots,\emptyset)
(\epsilon_1\leftrightarrow\epsilon_2)
\ee
where $i$ is the position of the non-empty diagram in the $N$-plet,
$Q_i(x) = \prod_{j\neq i} (x-a_j)(x-a_j+\epsilon)$
and these functions immediately reproduce the hypergeometric series
= the Fateev-Litvinov conformal block (\ref{hyse}):
\be
P_i(a_i) = \epsilon_1^N A_1\ldots A_N,\ \ \ \ \ \  \
Q_i(a_i) = {\epsilon_1^{N} B_1\ldots B_{N-1}\over -\epsilon_1\epsilon_2}
\label{PQvsAB}
\ee
for the chiral function.
However, there is a question: what guarantees that only
the chiral functions are non-vanishing?
As explained in \cite{mmNF}, this is a condition on polynomial $P$:
for given $i$,
\be
P_i(a_j) = 0 \ \ \ {\rm for} \ \ \ j\neq i
\ \ \ \ \ \ {\rm and} \ \ \ \ P_i(a_i+\epsilon_2) = 0
\label{Pcons}
\ee
These are, in fact, $N$ conditions on the coefficients
(or on the parameters $\mu_f$) of the polynomial
$P(a) = \prod_{f=1}^{N_f=2N}(a+\mu_f)$ of degree $2N$,
and the remaining $N$ parameters can be used to match the arbitrary
set of $A_1,\ldots,A_N$ in (\ref{PQvsAB}).
After that the $N-1$ parameters $a_j$ ($\sum_{j=1}^N a_j=0$)
can be adjusted to match $B_1,\ldots,B_{N-1}$.
Note that all these relations are {\it linear},
in accordance with the general AGT claim.
The parameters $\epsilon_1$ and $\epsilon_2$ are associated with the
central charge, and the common scale drops out from the Nekrasov
functions in the case of $N_f=2N$, as usual.

We now use the detailed description of the $N=3$ case
in this paper to show explicitly how these matchings work
in this particular case, and illustrate what (\ref{Pcons})
has to do with the maximal degeneracy condition imposed in
\cite{FLit}.
The simplest is to pick up formulas from the Conclusion, where they are collected
in a list.
For the sake of definiteness, we choose $i=3$ in (\ref{Pcons})
(the choice of $i$ corresponds to the choice of non-zero conformal block,
there are exactly $N$ of them).
Then, it is obvious  from (\ref{concavsal}) and (\ref{concmu})
that there is a natural choice of solution
to $P(a_1)=P(a_2)=0$:
\be
0 = (a_1+ \mu_3) = \frac{1}{\sqrt{3}}(\alpha-\alpha_2)
- (\beta-\beta_2) + S(\vec\alpha_1),\nn\\
0 = (a_2+\mu_2) = \frac{1}{\sqrt{3}}(\alpha-\alpha_2)
+ (\beta-\beta_2) + S(\vec\alpha_1)
\ee
i.e.
\be
\alpha = \alpha_2- \sqrt{3}S(\vec\alpha_1), \ \ \ \ \
\beta = \beta_2
\label{seru}
\ee
which restricts the intermediate momentum $\vec\alpha$ to
be a special combination of the two external momenta
$\vec\alpha_1$ and $\vec\alpha_2$.
The third condition $P(a_3+\epsilon_1)=0$ is then naturally
imposed as a constraint
\be
0 = \Big((a_3+\epsilon_1) +\mu_1\Big) =
-\frac{2}{\sqrt{3}}(\alpha-\alpha_2)+\epsilon_1 + S(\vec\alpha_1)
\ \stackrel{(\ref{seru})}{=} \ 3S(\vec\alpha_1) + \epsilon_1
\label{al1co}
\ee
which imposes an additional restriction on the
already-{\it special} momentum $\vec\alpha_1$, i.e.
fixes it completely.
With these choices, (\ref{seru}) and (\ref{al1co}) it is
guaranteed that the Nekrasov partition function is an $N=3$
hypergeometric series (\ref{hyse}) with
\be
P(a_3) = \prod_{f=1}^6 (a_3+\mu_f)
\ \stackrel{(\ref{seru})\&(\ref{al1co})}{=}\
(-\epsilon_1)\Big((\epsilon+\sqrt{3}\alpha_2)^2-\beta_2^2\Big)
\prod_{f=4}^6 (a_3+\mu_f)=\nn\\
=(-\epsilon_1)\Big((\epsilon+\sqrt{3}\alpha_2)^2-\beta_2^2\Big)
A_1A_2A_3
\ee
and
\be
a_{31}a_{32}(a_{31}+\epsilon)(a_{32}+\epsilon)
\ \stackrel{(\ref{seru})\&(\ref{al1co})}{=}\
\Big((\epsilon+\sqrt{3}\alpha_2)^2-\beta_2^2\Big)
(a_{31}+\epsilon)(a_{32}+\epsilon)=\nn\\=
\Big((\epsilon+\sqrt{3}\alpha_2)^2-\beta_2^2\Big)B_1B_2
\ee
i.e.
\be
A_s = a_3+\mu_{s+3} = \Big\{-{2\over\sqrt{3}}\alpha_2-{2\over 3}\epsilon+\mu_4,
-{2\over\sqrt{3}}\alpha_2-{2\over 3}\epsilon+\mu_5,
-{2\over\sqrt{3}}\alpha_2-{2\over 3}\epsilon+\mu_6 \Big\}, \ \ \
s= 1,2,3, \nn\\
B_s = a_{3s}+\epsilon = \Big\{\sqrt{3}\alpha_2-\beta_2,
\sqrt{3}\alpha_2+\beta_2 \Big\}, \ \ \
s= 1,2
\ee
with $\mu_{4,5,6}$ from (\ref{concmu}).

It remains to say that (\ref{al1co}) is exactly the
condition for $\vec\alpha_1$ to describe the maximally
degenerate state, and (\ref{seru}) is exactly the relevant
selection rule in the $sl(3)$ Toda model \cite{FLit}.
{\bf This completes the proof of AGT conjecture for the {\it generic}
$N=3$ hypergeometric conformal blocks.
The proof remains the same for arbitrary $N$,}  provided we
accept the straightforward generalization (\ref{genagtconc})
of (\ref{concavsal}) and (\ref{concmu}).
A similarly transparent proof for non-hypergeometric
blocks, however, remains to be found.

\section{Summary and conclusion}

Consideration in this paper leaves {\bf
little room for doubts that the AGT relations are true for the
Verma modules of generic groups}, {\it provided} appropriate
{\it speciality} (null-vector) constraints are imposed
on the modules.
Exact linear expressions for the Nekrasov
parameters for the $4$-point conformal block are\footnote{Note that
as compared with \cite{MMMagt}, our definition of
$\alpha$ is shifted: $\alpha\to\alpha+{\epsilon\over 2}$.}:
\be
SU(2):\ \ \ \ \ \ \ \ \ \ \ a_1=-a_2 = \alpha, \ \ \ \
\mu_{1,2} = \pm\alpha_2+S(\alpha_1), \ \ \ \
\mu_{3,4}=\pm\alpha_4+S(\alpha_3),\ \ \ \
\nu ={2S(\alpha_1)S(\alpha_2)\over\epsilon_1\epsilon_2}
\ee
for the $SU(2)$ case. There are two different choices for each of the two
shifts $S(\alpha_1)$ and $S(\alpha_3)$
(and thus a total of $4$ choices for the pair),
\be
S(\pm\alpha)={\epsilon\over 2}\pm\alpha
\ee
Similarly, for $SU(3)$
\be
\boxed{
a_1 = \frac{1}{\sqrt{3}}(\alpha-\beta\sqrt{3}),\ \ \
a_2 = \frac{1}{\sqrt{3}}(\alpha+\beta\sqrt{3}),\ \ \
a_3 = -\frac{2\alpha}{\sqrt{3}}
}
\label{concavsal}
\ee
\be
\boxed{
\begin{array}{ccc}
\mu_1=\frac{2}{\sqrt{3}}\alpha_2+S(\vec\alpha_1),
&& \mu_4=-\frac{2}{\sqrt{3}}\alpha_4+S(\vec\alpha_3), \\
\mu_2=-\frac{1}{\sqrt{3}}(\alpha_2+\beta_2\sqrt{3})+S(\vec\alpha_1),
&& \mu_5=\frac{1}{\sqrt{3}}(\alpha_4+\beta_4\sqrt{3})+S(\vec\alpha_3),\\
\mu_3=-\frac{1}{\sqrt{3}}(\alpha_2-\beta_2\sqrt{3})+S(\vec\alpha_1),
&& \mu_6=\frac{1}{\sqrt{3}}(\alpha_4-\beta_4\sqrt{3})+S(\vec\alpha_3)
\end{array}
}
\label{concmu}
\ee
\be
\boxed{
\nu = {3 S(\vec\alpha_1)S(\vec\alpha_3)\over\epsilon_1\epsilon_2}
}
\label{concnu}
\ee
There are three different choices for each of the two
shifts $S(\vec\alpha_1)$ and $S(\vec\alpha_3)$
(and thus a total of $9$ choices for the pair), associated with
the three possibilities to choose a {\it special} values\footnote{
We define these {\it special values} simply as zeroes
of the Kac determinant $K_1$, i.e. from the condition
$D\Delta^2=w^2$. Actual {\it special states}, satisfying
(\ref{Wzc}) do not obligatory exist for each of these values
(see, for example, the free-model example in \cite{MMMM}, where
only three out of six possible special {\it values} are actually
associated with special {\it states}). It turns out that for
this does not matter for AGT relation: it holds just for all
special {\it values}.
}
$\vec\alpha_1 = (\alpha_1,\beta_1)$ and
$\vec\alpha_3 = (\alpha_3,\beta_3)$:
\be
\boxed{
S(\alpha_1,-\epsilon/2) = -\frac{1}{\sqrt{3}}\alpha_1 + \frac{1}{2}\epsilon, \ \ \
S\Big(\alpha_1,-\alpha_1\sqrt{3} -\epsilon\Big)
= \frac{2}{\sqrt{3}}\alpha_1+\epsilon, \ \ \
S\Big(\alpha_1,\alpha_1\sqrt{3} - \epsilon\Big)
= \frac{2}{\sqrt{3}}\alpha_1}
\\
\boxed{
S(\alpha_3,-\epsilon/2) = \frac{1}{\sqrt{3}}\alpha_3 + \frac{1}{2}\epsilon, \ \ \
S\Big(\alpha_3,-\alpha_3\sqrt{3} - \epsilon\Big)
= -\frac{2}{\sqrt{3}}\alpha_3, \ \ \
S\Big(\alpha_3,\alpha_3\sqrt{3} - \epsilon\Big)
= -\frac{2}{\sqrt{3}}\alpha_3+\epsilon
}
\label{concshi}
\ee
If some other pair of external states is chosen to be {\it special},
formulas change according to the rule of projective transformation
of the conformal block. The resulting $\nu$ can be easily evaluated:
\be
1,3\leftrightarrow 2,4\ \ \ \ \ \ \nu\to {3S(\vec\alpha_2)S(\vec\alpha_4)
\over\epsilon_1\epsilon_2}+
\Delta_2+\Delta_4-\Delta_1-\Delta_3
\ee
(this rule can look more natural, if one mentions that
in the free field model there is an identity
$2\vec\alpha_2\vec\alpha_4 + \Delta_4+\Delta_2 - \Delta_1-\Delta_3
= 2\vec\alpha_1\vec\alpha_3$).
This same permutation with the corresponding change of $\nu\to
{2S(\alpha_2)S(\alpha_4)\over\epsilon_1\epsilon_2}-
\Delta_2-\Delta_4+\Delta_1+\Delta_3$
was discussed in the Liouville case
in \cite{MMMagt}.

With these values of parameters
\be\label{135}
\boxed{\boxed{
B^{(1)} = Z^{(1)} + \nu
\ \ \ \ \ \ \ {\rm and}\ \ \ \ \ \ \
B^{(2)} = Z^{(2)} + \nu Z^{(1)} +\frac{\nu(\nu+1)}{2},
}}
\ee
where $Z$ and $B$ are explicitly defined in (\ref{Nek2})
and (\ref{cbexp}) with substituted (\ref{SF1}), (\ref{SF2}) and
(\ref{vertices}).
Of many formulas in this paper these five are the only ones
used in the actual calculation at levels one and two (certainly, to check
(\ref{135}) one also needs the
identifications of parameters in (\ref{concavsal})-(\ref{concshi})).
All the rest is included in order to present the logic
and details of from-the-first-principles calculations very explicitly,
so that they can be straightforwardly validated and generalized.

Note that if formulas are rewritten in terms of the
fundamental weights vectors $\lambda_i$ from
(\ref{fundvec}), they
admit immediate generalization to arbitrary $N$.
The $sl(N)$ conformal Toda model has $W^{(N)}$ symmetry,
and the eigenvalues of its generators in $\alpha$-parametrization
are given by
\be
\Delta={\vec\epsilon^2-{\vec\alpha^2}\over\epsilon_1\epsilon_2},\ \ \ \
\ldots, \ \ \ \
w^{(N)}_{\vec\alpha} \sim \frac{1}{\sqrt{(-\epsilon_1\epsilon_2)^N}}
\prod_{i=1}^N (\vec\alpha\vec\lambda_i)
\ee
where
$\displaystyle{\vec\epsilon =
\epsilon\vec\rho}$
is directed along $\vec\rho$,
the half-sum of all positive roots (or the sum of the fundamental weights),
the central charge is $\displaystyle{c=(N-1)\left(1+N(N+1)
{\epsilon^2\over\epsilon_1\epsilon_2}\right)}$ and $\{\vec\lambda_i\}$
is a Weyl-symmetric set of $N$ minimal weights, see Fig.\ref{picroots}.
Then the AGT relation between the $sl(N)$ Toda
and $U(N)$ Nekrasov's functions should look like
\be
\boxed{
a_i =(\sqrt{2}\vec\alpha)\vec\lambda_i,\ \ \ \
\mu_i^+=-(\sqrt{2}\vec\alpha_2)\vec\lambda_i+S(\vec\alpha_1),\ \ \ \
\mu_i^-=(\sqrt{2}\vec\alpha_4)\vec\lambda_i+S(\vec\alpha_3),\ \ \ \
\nu  =\frac{NS(\vec\alpha_1)S(\vec\alpha_3)}{\epsilon_1\epsilon_2}
}
\label{genagtconc}
\ee
The $\sqrt{2}$ factors appear because with our normalization
conditions the free field primaries are $V_{\vec\alpha} =
e^{\sqrt{2}\vec\alpha\vec\phi}$. The notation $S$ refers simultaneously
to "shift" and "special".

\bigskip

Less technical is the general proof of the AGT relation.
A first step is made in \cite{mmNF} and s.\ref{FLpro}
of the present paper. It concerns the case, when
one of the two {\it special}
Verma modules in the 4-point conformal block is {\it further}
specialized by imposing two extra null-vector conditions which leads to
additional selection rules on the intermediate
state (enforced by the vanishing of the structure constants
$C_{\alpha_1,\vec\alpha_2,\vec\alpha}$, which are not
included into the definition of conformal block in the
present text).
In this particular case, the conformal block was explicitly
found in \cite{FLit} and it represents the generic hypergeometric
function $\phantom._NF_{N-1}$,
which has a straightforward character expansion.
As rightly anticipated in \cite{Wyl},
comparing it with Nekrasov formulas,
which in this case get contribution only from a
single set of "chiral" $N$-plets of Young diagrams, one obtains a
complete {\it proof} of the AGT relation in this particular case.
See \cite{mmNF} and s.\ref{FLpro} above.

The AGT relations can provide a new breath to abandoned areas
of conformal theory. The beautiful theory of $W$-algebras,
for example, may not attract enough attention, because
there was nothing to compare it with. Now {\it the equation gets
the other side}: whatever one obtains by hard calculations
in $W$ theory can be compared with the differently looking
formulas on the Nekrasov side. We demonstrated in this paper that
this, indeed, opens a possibility of validating and improving
$W$-algebra calculations.

As emphasized in \cite{MMMagt}, the AGT relation implies
that there are two different expansions of the same quantity
in characters (in sums over the Young diagrams):
the natural expansion of conformal blocks and the Nekrasov
sum over integer partitions. Already for $SU(2)$
they look absolutely different, like expansions
associated with free boson and free fermion formalisms.
The difference becomes even more pronounced for $N>2$:
it is enough to say that the conformal block expansion is in terms
of generalized Young diagrams ${\cal Y}$, labeling elements
of the $W_N$-algebra Verma modules, while the Nekrasov functions
are labeled by $N$-ples of the ordinary Young diagrams,
and there is no any {\it a priori} obvious relation between
the two.
It can deserve mentioning that the AGT relation does not link
contributions of individual diagrams on the l.h.s. and the r.h.s.,
only the entire sums over all diagrams of a given size (level).
Thus, at the same stage it can be useful to consider relations
between the conformal blocks of a given level and the original
integrals from \cite{Nek} which were afterwards decomposed into
sums of the Nekrasov functions. {\bf The AGT relation implies that these
integrals have two different complementary combinatorial
expansions} (we use the word combinatorial, like in \cite{MMMagt},
to emphasize that the character structure is not yet explicitly
revealed in these expansions).

A real mystery at the moment is what substitutes the AGT
relations in the case of generic Verma modules of the $W_N$ algebras.
The question can be asked in two opposite directions, that is,
about more {\it general} and more {\it special} cases: what happens to
generic {\it non-degenerate} modules for $N>2$ and
what happens to {\it stronger} degenerate modules
(say, associated with the Ashkin-Teller model \cite{ATM} and alike),
which have more sophisticated conformal blocks,
involving elliptic theta-constants already for $N=2$.
Answers to these questions can help to better understand
the group theory meaning and possible generalizations
of the Nekrasov functions.

The last but not least:
within the general framework, {\bf the AGT relation
between conformal theories and Nekrasov functions
looks very much like a "quantization"
of the well-known link \cite{GKMMM}
between Seiberg-Witten theory \cite{SW} and integrable systems}.
It is very important to understand the AGT relation
from this perspective and to compare it with the other
quantizations of the same link, like those of \cite{RG}-\cite{NS}.

\section*{Appendix: Normalization conventions \label{norcft}}

We use somewhat non-standard normalizations,
adjusted to maximally simplify the $W$-algebra formulas.
The price for this is certain deviations from
choices made in some other papers.
We illustrate our conventions with the example of free field
theory.

The freedom in the choice of normalization prescriptions in CFT
can be described as follows.
First, one can choose the coefficient in the action,
i.e. normalization of the free field, or, equivalently, the coefficient
in the propagator, we call it $\kappa$.
If the stress tensor is defined as a generator of (local) translations,
i.e. has the standard operator product expansion with the term
$T(z_1)V(z_2) = \ldots + z_{12}^{-1}\p V(z_2) + \ldots$
with the {\it unit} coefficient in front of the $V$-derivative term,
then, $\kappa^{-1}$ appears as the coefficient in $T$, thus,
$\kappa$ can be alternatively considered as a freedom in
normalization of $T=W^{(2)}$.
Similarly, to each operator $W^{(k)}$ (there are $N-1$ of them
in the theory with $N-1$ free fields) can be ascribed
an arbitrary new coefficient $\kappa_k$, i.e. there is no way to tie
the normalization of $W^{(k)}$'s with $k>2$ to that of the propagator,
and there is no {\it canonical} way to fix normalizations of
some structure constants of the $W$-algebra.
When the central charge $c$ deviates from $N-1$, there is a new
arbitrariness in the choice of the normalization parameter
in the deformation term $\p^2\phi$ in $T$: in fact,
it can be considered as a freedom to normalize $W^{(1)}$.
Second, in addition to these $N$ free normalizations, different
papers define differently the parameter $\lambda$ in the conformal dimensions:
$\Delta_{\vec\alpha} = \Delta\big(e^{\vec\alpha\phi/\lambda}\big)$.
So far, fortunately the same $\lambda$ always appeared in the
definition of other eigenvalues:
$w_{\vec\alpha} = w\big(e^{\vec\alpha\phi/\lambda}\big)$ and so on.

More explicitly,
if the operator product expansion of the free fields is defined as
\be
\p\phi_i(z)\p\phi_j(0) = \kappa_2\frac{\delta_{ij}}{z^2} + \ldots
\ee
then the stress tensor
\be
T(z) = \frac{1}{2\kappa_2}\sum_{i=1}^{N-1}(\p\phi_i)^2
+ \kappa_1\sum_{i=1}^{N-1} Q_i\p^2\phi_i
= \frac{1}{2\kappa_2}(\p\vec\phi)^2 + \kappa_1\vec Q\p^2\vec\phi
\ee
satisfies
\be
T(z)T(0) = \frac{c/2}{z^4} + \frac{2T(0)}{z^2}
+ \frac{\p T(0)}{z}  + \ldots
\ee
with
\be
c = N-1 - 12\kappa_2\kappa_1^2\vec Q^2,
\ee
while the primary field
\be
V_{\vec\alpha}
= \ :\exp\left(\frac{\vec\alpha\vec\phi}{\lambda}\right):
\ee
satisfies
\be
T(z) V_{\vec\alpha}(0) =
\frac{\Delta_{\vec\alpha}}{z^2} V_{\vec\alpha}(0)
+ \frac{1}{z}L_{-1}V_{\vec\alpha}(0) + \ldots
\ee
with the dimension
\be
\Delta_{\vec\alpha} = \frac{\kappa_2}{\lambda^2}\cdot
\frac{\vec\alpha(\vec\alpha - 2\lambda\kappa_1\vec Q)}{2}
\ee
The free field theory selection rule for the
non-vanishing correlator of exponentials is
\be
\sum_i \vec\alpha_i = 2\lambda\kappa_1\vec Q = 2\lambda\varepsilon\vec\rho
\ee
It is inspired by zero-mode integration within the functional integral
approach, or by conditions like
\be
\langle V_{\vec\alpha_1}(z_1) V_{\vec\alpha_2}(z_2)\rangle \sim
(z_1-z_2)^{\kappa_2/\lambda^2} = \frac{1}{(z_1-z_2)^{\Delta_{\vec\alpha_1}
+ \Delta_{\vec\alpha_2}}}\ \  \ \Longrightarrow \ \ \
\Delta_{\vec\alpha_1}=\Delta_{\vec\alpha_2}
= -\frac{\kappa_2}{2\lambda^2}\,\vec\alpha_1\vec\alpha_2
\ee
within CFT itself.

Usually in CFT with $N\neq 2$ \cite{fref}
one chooses $\vec Q$ to be directed along the vector
$\vec\rho$, the half-sum of all positive roots of $sl(N)$
or, which is the same, the sum of all fundamental weights,
$|\vec\rho\,|^2 = \frac{N(N^2-1)}{12}$
(for non-simply-laced groups
$\vec Q \sim \sqrt{-\epsilon_1/\epsilon_2}\,\vec\rho +
\sqrt{-\epsilon_2/\epsilon_1}\,\vec\rho\, \check{\phantom.}$).
The proportionality coefficient, however,
is not canonically fixed and remains an arbitrary
normalization parameter.
The apparent problem with this convention, uniform and convenient
for all groups, is that for $SU(2)$ $\rho = 1/\sqrt{2}$,
therefore, one rarely uses it in considerations of the single field
case, since it makes $k_1$ to contain $\sqrt{2}$.
One can continue this "natural" $N=2$ convention to all $N$,
using coordinates where $\vec\rho\,\vec\phi/|\rho| = \phi_{N-1}$,
and this choice also is present in the literature. Anyhow,
\be
\boxed{
\ \ {\rm if} \ \ \ \ \kappa_1\vec Q = \varepsilon\vec\rho, \ \ \ \
{\rm then} \ \ \ \ c = (N-1)\Big(1 - N(N+1)\kappa_2\varepsilon^2\Big)\ \ \ \
{\rm and} \ \ \ \ \Delta_{\vec\alpha} =
\frac{\kappa_2}{2\lambda^2}\,\vec\alpha
\big(\vec\alpha-2\lambda\kappa_1\vec Q\big)\
}
\ee
At the same time,
the central charge of the $sl(N)$ Toda model, expressed by the AGT relation
through the Nekrasov parameters
$\epsilon=\epsilon_1+\epsilon_2$, is equal to
\be
\boxed{
c_N = (N-1)\left(1 + N(N+1)\frac{\epsilon^2}{\epsilon_1\epsilon_2}\right)
= \left\{\begin{array}{ccc}
1 + \frac{6\epsilon^2}{\epsilon_1\epsilon_2} &\ {\rm for}\ & N=2,\\
&&\\
2 + \frac{24\epsilon^2}{\epsilon_1\epsilon_2} & {\rm for} &N=3,\\
\ldots &&
\end{array}
\right.
}
\ee
i.e. in this notation
\be
\varepsilon = \frac{1}{\sqrt{\kappa_2}}
\cdot\frac{\epsilon}{\sqrt{-\epsilon_1\epsilon_2}}
\ee
Since the common rescaling of all parameters in the Nekrasov functions
with $N_f=2N$ does not affect the answer, one can easily choose
$\epsilon_2=-1/\epsilon_1$, as it was done, for example, in
\cite{AGT} and \cite{mmNF}.
The following table contains the comparison of notations in
different papers:

\bigskip

\centerline{$
\begin{array}{|c|c|c|c|c|c|c|c|c|c|}
\hline
&&&&{\rm selecton}&&&&&\\
{\rm paper} & $N$
& {\rm central\ charge}&{\rm dimension}
&{\rm rule:} &  {\rm exponential}
 &1/\lambda & \varepsilon/Q = \kappa_1/|\rho| & \kappa_1\vec Q &\kappa_2\\
&&&&\sum_i \vec\alpha_i = &{\rm primary}&&&
&\\
\hline
&&&&&&&&&\\
\cite{Wyl}&2&1+6Q^2&\alpha(Q-\alpha)&Q&:e^{2\alpha\phi}:&2&\sqrt{2}&Q&-1/2\\
&&&&&&&&&\\
\hline
&&&&&&&&&\\
\cite{Wyl}&N&(N-1)\Big(1+
&\frac{\vec\alpha(2\vec Q -\vec\alpha)}{2}
&2\vec Q&:e^{\vec\alpha\vec\phi}:&1&\frac{1}{|\rho|}=
\sqrt{\frac{12}{N(N^2-1)}}&\vec Q&-1\\
&&+ N(N+1)\varepsilon^2\Big)&&&&&&&\\
\hline
&&&&&&&&&\\
\cite{MMMagt}&2&1+\frac{6Q^2}{\epsilon_1\epsilon_2}
&\frac{\alpha(Q-\alpha)}{\epsilon_1\epsilon_2}&Q
&:e^{\alpha\phi}:&1&\frac{1}{2\sqrt{2}}
&\frac{1}{2}Q&\frac{2}{\epsilon_1\epsilon_2}\\
&&&&&&&&&\\
\hline
&&&&&&&&&\\
\cite{MMMM}&2& 1-12Q^2
&\frac{\alpha(\alpha-2Q)}{2}&2Q
&:e^{\alpha\phi}:&1&|\rho_{SU(2)}|^{-1}=\sqrt{2}&Q& 1\\
&&&&&&&&&\\
\hline
&&&&&&&&&\\
\cite{MMMM}&3&2-12Q^2
&\frac{\alpha^2+\beta(\beta-2Q)}{2}&(0,2Q)
&:e^{\alpha\phi_1+\beta\phi_2}:&1
&|\rho_{SU(3)}|^{-1}=\frac{1}{\sqrt{2}}&(0,Q)&1\\
&&&&&&&&&\\
\hline
&&&&&&&&&\\
{\rm present}&3&2-24Q^2&\alpha^2+\tilde\beta^2-Q^2&(0,2Q)
&:e^{\sqrt{2}(\alpha\phi_1+\beta\phi_2)}:&\sqrt{2}&1
&(0,\sqrt{2}Q)&1\\
{\rm paper}&&&\tilde\beta = \beta-Q&&&&&&\\
&&&&&&&&&\\
\hline
&&&&&&&&&\\
\cite{mmNF}\ {\rm and}&&&&&&&&&\\
{\rm present}&N&(N-1)\Big( 1 +
&\frac{\vec\alpha(\vec\alpha-2\vec Q)}{-\epsilon_1\epsilon_2}
&2\vec Q
&:e^{\sqrt{2}\vec\alpha\vec\phi}:
&\sqrt{2}&\frac{\sqrt{2}}{|\vec\rho\,|}
&\sqrt{2}\vec Q&\frac{1}{-\epsilon_1\epsilon_2}\\
{\rm paper}
&&+\frac{N(N+1)\varepsilon^2}{\epsilon_1\epsilon_2}\Big)
&=\frac{\overrightarrow{\tilde\alpha}^2-Q^2}{-\epsilon_1\epsilon_2}&&&&&&\\
&&&&&&&&&\\
\hline
\end{array}
$}

\bigskip

In sample free field considerations
with $N=3$ in this paper we imply that
$V_{\vec\alpha} = e^{\sqrt{2}(\alpha\phi_1 + \beta\phi_2)}$.
Thus, it is natural to normalize
the $W^{(3)}$ operator to be
\be
W^{(3)} = \frac{1}{2\sqrt{2}}
\Big((\p\phi_1)^3 - 3\p\phi_1(\p\phi_2)^2\Big)
\ee
so that
\be
W^{(3)}(z) V_{\vec\alpha}(0) =
\frac{w_{\vec\alpha}}{z^3} V_{\vec\alpha}(0)
+ \frac{1}{z^2}W_{-1}V_{\vec\alpha}(0)
+ \frac{1}{z}W_{-2}V_{\vec\alpha}(0) \ldots
\ee
with\footnote{In fact, it is
$w_{\vec\alpha} = {(-\epsilon_1\epsilon_2)^{-3/2}}\cdot
\alpha(\alpha^2-3\beta^2)$, but we systematically suppress
powers of $(-\epsilon_1\epsilon_2)$ in most formulas
to make them more readable.
One should remember they are actually there.
}
\be
w_{\vec\alpha} =
\alpha(\alpha^2-3\beta^2)
\ee
and
\be
W_{-1}V_{\vec\alpha} =\ : \frac{3}{\sqrt{2}}
\Big((\alpha^2-\beta^2)\p\phi_1 - 2\alpha\beta\p\phi_2\Big)V_{\vec\alpha}:
\label{W-1V}
\ee
Also, of course,
\be
L_{-1}V_{\vec\alpha} =
\ :\sqrt{2}(\alpha\p\phi_1 + \beta\p\phi_2)V_{\vec\alpha}:\
= \p V_{\vec\alpha}
\label{L-1V}
\ee
The {\it special} primaries of the $W_3$ algebra generate the
Verma modules with a null-vector at the first level,
with $W_{-1}V_{\vec\alpha} = \zeta L_{-1}V_{\vec\alpha}$
for some $\lambda$. From (\ref{L-1V}) and (\ref{W-1V}) it follows
that
\be
3(\alpha^2-\beta^2) = 2\zeta\alpha, \nn \\
-6\alpha\beta = 2\zeta\beta
\ee
i.e. $\zeta = -3\alpha$ and $\beta^2=3\alpha^2$, or $\zeta=3/2\alpha$ and $\beta=0$.
Note that, under these relations,
$\zeta = 3w_{\vec\alpha}/2\Delta_{\vec\alpha}$,
so that the null-vector condition is exactly (\ref{Wzc}) and
the Kac determinant $v^2=\Delta^3-w^2
= \Big(\beta(\beta^2-3\alpha^2)\Big)^2$ vanishes.

The main peculiarity of our normalization conventions
is that the dimension $\Delta = \vec\alpha^2$ does not contain
a minus sign, in variance from \cite{AGT,Wyl,MMMagt}:
in dealing with $W$-algebras this eliminates unnecessary
imaginary units.
Factors of $2$ are eliminated from
$\Delta$ and $w$ to make the central relation (\ref{Deltaw})
as simple as possible. The price for this is the roots of $2$
in the primary exponentials and $W$ operators.
Roots of $3$, however, are unavoidable,
they come from the roots and weights of the underlying
$sl(3)$ algebra.

\section*{Acknowledgements}

We are indebted for hospitality and support to Prof.T.Tomaras and
the Institute of Theoretical and Computational Physics of
University of Crete, where part of this work was done.

The work was partly supported by Russian Federal Nuclear Energy
Agency and by RFBR grants 07-02-00878 (A.Mir.),
and 07-02-00645 (A.Mor.).
The work was also partly supported by grant FP7-REGPOT-1 (Crete HEP Cosmo 228644),
by joint grants 09-02-90493-Ukr,
09-02-93105-CNRSL, 09-01-92440-CE, 09-02-91005-ANF, INTERREG IIIA
(Greece-Cyprus) and by Russian President's Grant
of Support for the Scientific Schools NSh-3035.2008.2.

\end{document}